\documentclass[11pt, twoside]{amsart}

\usepackage[utf8]{inputenc}
\usepackage[T1]{fontenc}
\usepackage{amssymb,amsmath,amstext}
\usepackage{hyperref, graphicx}
\usepackage{stackrel}

\newtheorem{theor}{Theorem}[section]
\newtheorem{lem}[theor]{Lemma}
\newtheorem{defin}[theor]{Definition}

\newtheorem{prop}[theor]{Proposition}

\newtheorem{exam}[theor]{Example}

\newtheorem{cor}[theor]{Corollary}
\newtheorem{rem}[theor]{Remark}

\newtheorem{assump}[theor]{Assumption}

\numberwithin{equation}{section}

\newcommand{\mr}{\mathrm}

\newcommand{\es}{\emptyset}

\newcommand{\uhr}{\upharpoonright}

\newcommand{\nts}{\negthickspace}
\newcommand{\uhrc}{\nts \upharpoonright \nts}

\newcommand{\mcA}{\mathcal{A}}
\newcommand{\mcB}{\mathcal{B}}
\newcommand{\mcC}{\mathcal{C}}

\newcommand{\mcG}{\mathcal{G}}

\newcommand{\mff}{\mathfrak{f}}
\newcommand{\mfg}{\mathfrak{g}}

\newcommand{\mbW}{\mathbf{W}}
\newcommand{\mbX}{\mathbf{X}}
\newcommand{\mbY}{\mathbf{Y}}

\newcommand{\mbbG}{\mathbb{G}}
\newcommand{\mbbP}{\mathbb{P}}
\newcommand{\mbbN}{\mathbb{N}}
\newcommand{\mbbR}{\mathbb{R}}

\newcommand{\rng}{\mathrm{rng}}

\title[Asymptotic elimination of aggregation functions]
{Asymptotic elimination of partially continuous aggregation functions in directed graphical models}

\author{Vera Koponen and
Felix  Weitkämper}

\address{Vera Koponen, Department of Mathematics, Uppsala University, Sweden.}

\email{vera.koponen@math.uu.se}

\address{Felix Weitkämper, Institut für Informatik,
Ludwig-Maximilians-Universität München, Munich, Germany.}

\email{felix.weitkaemper@lmu.de}

\date{28 June 2023}

\begin{document}

\begin{abstract}
In statistical relational artificial intelligence, a branch of AI and machine learning which combines the logical and statistical schools of AI, 
one uses the concept of a
{\em para\-metrized probabilistic graphical model (PPGM)}
to model (conditional) dependencies between random variables and
to make probabilistic inferences about events on a space of ``possible worlds''.
The set of possible worlds with underlying domain $D$ (a set of objects) can be represented by the set $\mathbf{W}_D$
of all first-order structures (for a suitable signature) with domain $D$.
Using a formal logic we can describe events on $\mathbf{W}_D$. 
By combining a logic and a PPGM we can also define a probability distribution $\mathbb{P}_D$ on $\mathbf{W}_D$ and use it to
compute the probability of an event.
We consider a logic, denoted $PLA$, with truth values in the unit interval, which uses aggregation functions instead of quantifiers.
This is motivated by the fact that aggregation functions such as arithmetic mean, geometric mean, maximum and minimum are
important tools in analysis of data.

However, we face the problem of computational efficiency, and this problem is an obstacle to the wider use of 
methods from Statistical Relational AI in practical applications.
The brute force way of computing, for $S \subseteq [0, 1]$ and a $PLA$-sentence $\varphi$, the probability that
the value of $\varphi$ belongs to $S$ 
needs an amount of time which grows exponentially in the size of $D$.
We address this problem by proving that the described probability will, under certain assumptions on the PPGM
and the sentence $\varphi$, converge as the size of $D$ tends to infinity.
The convergence result is obtained by showing that every formula $\varphi(x_1, \ldots, x_k)$ which contains only ``admissible''
aggregation functions (e.g.\ arithmetic and geometric mean, max and min) is asymptotically equivalent 
to a formula $\psi(x_1, \ldots, x_k)$ without aggregation functions. This means that for every $\varepsilon > 0$ the probability
that, for some parameters $a_1, \ldots, a_k \in D$, the values of $\varphi(a_1, \ldots, a_k)$ and 
$\psi(a_1, \ldots, a_k)$ differ by more than $\varepsilon$ approaches 0 as the domain size tends to infinity.
The proof provides a method for finding such $\psi$, given only $\phi$ and the PPGM as input.
\end{abstract}

\maketitle

\section{Introduction}

\subsection{Aggregation functions}

Aggregation functions (also called aggregate functions or combination functions) are an important tool in analysis of data.
Such functions take a sequence of numbers (or more generally, some number of sequences of numbers) and return
a number.
Here we will only consider aggregation functions whose value does not depend on the order of the numbers in the sequence.
Typical examples include the arithmetic mean, the geometric mean, and the maximum of the numbers in the sequence. 
Moreover, we consider only sequences of numbers in the unit interval $[0, 1]$ and aggregation functions with values in $[0, 1]$.
In our context this is natural because the numbers we consider can be viewed as probabilities (or relative frequences) and the
 logic that we will consider has truth values in $[0, 1]$.
But one can also think of numbers in the unit interval as being the ``normalized'' versions of numbers in $[0, a]$
for some positive $a \in \mbbR$.
As usual $[0, 1]^n$ denotes the Cartesian product of $n$ intervals $[0, 1]$ and we let 
$[0, 1]^{<\omega} = \bigcup_{n=1}^\infty [0, 1]^n$.
Now we can as well make precise what we mean by an aggregation function in this article:

\begin{defin}\label{definition of aggregation function}{\rm
Let $F : \big([0, 1]^{<\omega}\big)^k \to [0, 1]$, so $F$ takes $k$ sequences (not necessarily of the same length) as input.
We call $F$ an {\em aggregation function} if $F$ is symmetric in the sense that if $\bar{r}_1, \ldots, \bar{r}_k  \in [0, 1]^{<\omega}$ 
and for each $i = 1, \ldots, k$, $\bar{\rho}_i$ is an arbitrary reordering of the entries of $\bar{r}_i$, then
$F(\bar{\rho}_1, \ldots, \bar{\rho}_k) = F(\bar{r}_1, \ldots, \bar{r}_k)$.
}\end{defin}

\begin{exam}\label{remark on common aggregation functions} {\bf (Common aggregation functions)} {\rm
The aggregation functions listed below are common when analyzing data. 
For $\bar{r} = (r_1, \ldots, r_n) \in [0, 1]^{<\omega}$, define
\begin{enumerate}
\item $\max(\bar{r})$ to be the {\em maximum} of all $r_i$,
\item $\min(\bar{r})$ to be the {\em minimum} of all $r_i$,
\item $\mr{am}(\bar{r}) = (r_1 + \ldots + r_n)/n$, so `am' is the {\em arithmetic mean}.
\item $\mr{gm}(\bar{r}) = \big(\prod_{i=1}^n r_i\big)^{(1/n)}$, so `gm' is the {\em geometric mean}.
\item $\text{noisy-or}(\bar{r}) = 1 - \prod_{i=1}^n (1 - r_i)$.
\end{enumerate}
}\end{exam}

\subsection{Logic}

We will study a {\em probability logic with aggregation functions}, abbreviated $PLA$.
Since the output of an aggregation function may be any number in the unit interval it follows that $PLA$ will be a many valued logic with
truth values in the unit interval.
Since the aggregation functions max and min can be used to express existential and universal quantification, respectively,
it follows that the expressive power of $PLA$ exceeds that of first-order logic.
Examples of the expressivity of $PLA$ are given in Section~\ref{Discussion}.
For example we show that every stage of the SimRank \cite{JW} can be expressed by a $PLA$-formula.

The syntax of $PLA$ (Definition~\ref{syntax of PLA}) is similar to the probability logic studied by Jaeger~\cite{Jae98a}, 
but we use the semantics (Definition~\ref{semantics of PLA})
of Lukasiewicz logic for the propositional connectives $\neg$, $\to$, $\vee$ and $\wedge$.
We make this choice because we want the truth value of, for example, $\varphi \to \psi$ to vary continuously 
as the truth values of $\varphi$ and $\psi$ vary.
When formulas take the truth values 0 or 1 the semantics of 
Lukasiewicz coincides with the common semantics of the mentioned connectives.
For a concise introduction to Lukasiewicz logic see e.g. \cite[Section~11.2]{Ber}, or see the original source \cite{LT}.

\subsection{Probability distributions and parametrized probabilistic gra\-phi\-cal mod\-els}

Formulas of $PLA$ are evaluated in finite structures, which can be thought of as ``possible worlds''.
In particular, given a formula we are interested in 
the probability that 
this formula takes a particular (truth) value, or a value in a given interval, when interpreted 
in a random possible world with a fixed domain.
The problem formulation assumes that a probability distribution is given on the set of possible worlds 
with a fixed domain. 
We fix an arbitrary finite first-order signature $\sigma$ with only relation symbols. In practice the signature is
determined by the context.
We assume that the domain is $[n] = \{1, 2, \ldots, n\}$ for some positive integer $n$.
Let $\mbW_n$ denote the set of all $\sigma$-structures (in the usual sense of first-order logic) with domain $[n]$.
There are many ways to define a probability distribution on $\mbW_n$.
Since our aim is to obtain results that are useful within the context of {\em statistical relational AI},
a subfield of AI and machine learning, we consider a probability  distribution on $\mbW_n$ which is
determined by a so-called {\em parametrized (or lifted) probabilistic graphical model (PPGM)}.
(For background on statistical relational learning and probabilistic graphical models see e.g. \cite{BK, DKNP, GT, KMG}.)
A PPGM is determined partly by a (directed or undirected) graph, the vertices of which are so-called {\em parametrized random variables},
but can also be seen as (usually) atomic first-order formulas. The (conditional) dependencies between the random variables
are expressed by the edges of the graph. 
For every parametrized random variable of a directed PPGM 
the conditional probability of it taking a given value can be computed from the values of its parents.
The formalization of a PPGM used in this article, first considered in \cite{Kop20}, is a {\em lifted Bayesian network}
in the sense of
Definition~\ref{definition of BN}
below which uses {\em conditional probability logic (CPL)} 
(Definition~\ref{definition of CPL}) to express ``threshold conditions''.
Informally speaking, a lifted Bayesian network assigns a probability to an atomic formula $R(\bar{x})$
by saying that it has probability $\alpha_i$ if a condition $\chi_i(\bar{x})$ holds where $\chi_i(\bar{x})$ may
not use $R$
but may use the usual syntactic constructs of first-order logic and constructs with the following meaning:
The relative frequency of $\bar{y}$ satisfying $\varphi_1(\bar{x}, \bar{y})$ conditioned on
$\bar{y}$ satisfying $\varphi_2(\bar{x}, \bar{y})$ is at least as large as
the relative frequency $\bar{y}$ satisfying $\varphi_3(\bar{x}, \bar{y})$ conditioned on $\bar{y}$
satisfying $\varphi_4(\bar{x}, \bar{y})$.
Thus lifted Bayesian networks are well suited for expressing probabilities that change if a threshold
(in terms of a relative frequency) is passed
but which stay fixed between the thresholds.
The discussion in 
Section~\ref{Discussion}, including Example~\ref{example of LBN}, hints to what kind of distributions can be expressed and their relevance.

\subsection{Problems and results}

With a lifted Bayesian network and a description of how they induce
a probability distribution $\mbbP_n$ on $\mbW_n$ 
(Definition~\ref{the probability distribution induced by an LBN})
we have the framework for asking for a PLA-formula $\varphi(\bar{x})$ and $\bar{a} \in [n]^k$ 
(where  $k$ is the length of the sequence of variables $\bar{x}$):
What is the probability that $\varphi(\bar{a})$ has a given value (or that its value belongs to a given interval)?
The ``brute force'' method to compute this probability is to compute the value of $\varphi(\bar{a})$ in all
structures in $\mbW_n$ and then add the probabilities of those structures where $\varphi(\bar{a})$ has the given value(s).
Needless to say, this approach is in general extremely inefficient for large $n$.
Ideally, we would like to have a way of computing, or at least approximating, 
the probability that $\varphi(\bar{a})$ takes a certain value which is independent
of the size $n$ of the domain.
If $\varphi(\bar{x})$ has no aggregation functions then the value of 
$\varphi(\bar{a})$ depends only on which atomic formulas the sequence $\bar{a}$ satisfies;
it follows that, for any interval $I \subseteq [0, 1]$,
the probability that $\varphi(\bar{a})$ belongs to $I$ can be computed 
by using {\em only} the lifted Bayesian network and $\varphi$.

Thus, from a computational perspective, we may wish that for every formula $\varphi(\bar{x})$ of $PLA$
we could find a formula $\psi(\bar{x})$ without aggregation functions such that
with a probability approaching 1 as $n\to\infty$
$\varphi(\bar{a})$ and $\psi(\bar{a})$ have the same value.
(In the context of a 2-valued logic with quantifiers instead of aggregation functions 
such a result would be called ``almost sure elimination of quantifiers''.)
However, it is not difficult to construct a formula $\varphi(\bar{x})$
such that (even under the assumptions of the main theorems of this article)
there is no $\psi(\bar{x})$ without aggregation functions such that, almost surely, $\varphi(\bar{a})$ and $\psi(\bar{a})$
have the same value for all $\bar{a}$.
Therefore we consider the notion of {\em asymptotic equivalence}.
Two formulas $\varphi(\bar{x})$ and $\psi(\bar{x})$
are asymptotically equivalent 
(Definition~\ref{definition of asymptotically equivalent formulas})
if for every $\varepsilon > 0$, the probability that there is $\bar{a}$ such that
the difference between the values of $\varphi(\bar{a})$ and $\psi(\bar{a})$ is larger than $\varepsilon$ tends to 0 as $n\to\infty$.
Note that in a two-valued logic asymptotic equivalence and almost sure equivalence coincide.

Due to the wide variety of aggregation functions and probability distributions we do not expect to be able to
``asymptotically eliminate'' all possible aggregation functions for all possible probability distributions.
Our main result,
Theorem~\ref{elimination of aggregation functions for an LBN},
says that if a lifted Bayesian network has the property that all of its aggregation formulas
are {\em noncritical} (in a sense to be made precise),
then every formula of $PLA$ with only {\em admissible} aggregation functions is asymptotically equivalent to 
a formula without aggregation functions, with respect to probability distributions induced by the lifted Bayesian network. 
The condition that an aggregation function is admissible means, very roughly, that it behaves in a uniformly continuous
way within certain restricted contexts.
As explained by Remark~\ref{Computing an asymptotically equivalent formula without aggregation functions},
the asymptotically equivalent formula without aggregation functions can be computed from the original formula by 
using only the lifted Bayesian network that induces the probability distribution. 
The result about asymptotic elimination of admissible aggregation functions implies that 
for every $PLA$-formula with only admissible aggregation functions (which include max, min, arithmetic and geometric means)
the probability that it is satisfied (by a random tuple of parameters) converges as the domain size tends to infinity.
To the best of our knowledge this is the first convergence law of a logic with truth values in 
the unit interval $[0, 1]$ and which can express all properties that are expressible in first-order logic.

\subsection{Related work}\label{Related work}

Already in 1998 Jaeger \cite{Jae98a} proved a convergence result for first-order formulas in a context where the 
probability distribution was determined by a {\em relational Bayesian network} that uses only
{\em exponentially convergent} aggregation functions.
The logic $PLA$ that we will define is much like Jaeger's {\em probability logic} \cite{Jae98a},
but we will use $PLA$ for defining queries, while Jaeger's use of probability logic in \cite{Jae98a} is to define probability distributions
(via relational Bayesian networks).

More recently, Koponen \cite{Kop20} proved ``almost sure elimination of conditional
probability quantiers'' and (as a by product) a zero-one law for conditional probability logic, which
is a two valued logic that extends first-order logic, in a context where the distribution was determined by a
lifted Bayesian network.
Quite recently, Grädel et al. \cite{Gra} proved convergence laws for first-order logic with 
semiring semantics.

Besides the above results there has recently been a growing interest in the AI community
in investigating the effect of increasing domain sizes on probabilistic inference in various contexts.
Very limited convergence results with respect to logical expressibility, 
covering only Boolean combinations of atomic formulas, have been obtained for 
{\em (domain-aware) Markov logic networks} and 
{\em relational logistic regression networks} by Poole et al. \cite{Poole14} and  Mittal et al. \cite{Mittal19}.
Weitkämper \cite{Wei21a} and \cite{Wei21c} showed that domain-aware relational logistic regression networks and, more generally, functional lifted Bayesian networks are asymptotically equivalent to aggregation-free networks. However, they only allow for non-nested dependencies on relative frequencies rather than allowing for a choice of aggregation function. Weitkämper \cite{Wei21b} shows asymptotic quantifier elimination for probabilistic logic programming, which only supports the noisy-or combination function.  

Jaeger's work in \cite{Jae98a} considers {\em exponentially convergent} aggregation functions in the probability formulas
used to define probabilities in  relational Bayesian networks. 
Although his notion of exponentially convergent aggregation function is similar in spirit to our
Definition~\ref{alternative definition of admissible function} of {\em admissible}
aggregation function, neither of the notions implies the other.
Indeed, noisy-or is exponentially convergent but not admissible, 
while the arithmetic mean function is admissible but not exponentially convergent.


\subsection{Organization}

Section~\ref{Preliminaries} clarifies some basic terminology and notation.
Section~\ref{Probability logic with aggregation functions} defines the syntax and semantics of
$PLA$ and derives a couple of basic properties of $PLA$.
Section~\ref{PGM} introduces the reader to asymptotic equivalence of formulas,
conditional probability logic and lifted Bayesian networks, and the way they induce a probability distribution.
In Section~\ref{Discussion} we discuss the expressivity of lifted Bayesian networks
and of $PLA$, including concrete examples.
In Section~\ref{Admissibility and the main results} the notion of {\em admissible} aggregation function is defined.
It is proved that arithmetic mean, geometric mean, maximum, minimum and conditional arithmetic mean
are admissible aggregation functions and the main result, 
Theorem~\ref{elimination of aggregation functions for an LBN},
and its corollary about convergence are stated.
Section~\ref{asymptotic elimination of aggregation functions} contains the proof of 
Theorem~\ref{elimination of aggregation functions for an LBN}.

\section{Preliminaries}\label{Preliminaries}

\noindent
We use more or less standard notation and terminology within the field of finite model theory; see e.g. \cite{Lib}.
The letter $\sigma$ (or $\sigma'$) will always denote a finite relational signature (vocabulary). 
By saying that $\sigma$ is {\em finite and relational} we mean that $\sigma$ is finite and contains only relation symbols.
We use the expression {\em $\sigma$-structure} in the sense of first-order logic and such structures are denoted 
by calligraphic letters $\mcA, \mcB, \mcC, \ldots$,
possibly with super- or subscripts. 
If $\mcA$ is a $\sigma$-structure and $\sigma' \subset \sigma$ then $\mcA \uhrc \sigma'$ denotes the {\em reduct} of $\mcA$
to the (sub)signature $\sigma'$.
The {\em domain (universe)} of a structure $\mcA$ will be denoted by the corresponding noncalligraphic letter $A$. 
Often the domain $A$ will be the 
set $[n] = \{1, \ldots, n\}$ for some integer $n \in \mbbN^+$ where $\mbbN^+$ denotes the set of all positive integers and
$\mbbN$ denotes the set of all nonnegative integers.
The {\em cardinality} of a set $A$ is denoted by $|A|$.
Finite sequences (tuples) of elements are denoted by $\bar{a}$ for some noncapital letter $a$.
The {\em length} of a sequence $\bar{a}$ is denoted $|\bar{a}|$.
For two sequences $\bar{a}$ and $\bar{b}$ their concatenation is denoted $\bar{a}\bar{b}$.
The set of all elements that occur in a sequence $\bar{a}$ is called its {\em range} and is denoted $\mr{rng}(\bar{a})$.
For a set $A$ and integer $k > 0$, $A^k$ denotes the set of all $k$-tuples (sequences of length $k$) of elements from $A$
and $A^{<\omega} = \bigcup_{k \in \mbbN^+}A^k$.

The letters $x, y, z$ (possibly with indices) will almost always denote formal logical variables. 
The expressions $\bar{x}, \bar{y}, \bar{z}$ will denote finite sequences of {\em distinct} variables although this 
assumption may be repeated sometimes. However if $\bar{a}$ denotes a sequence of some other kind, a sequence of reals for example,
then we allow repetitions of the same element in the sequence.
Formulas of a formal logic are usually denoted $\varphi$, $\psi$, $\chi$ or $\theta$.
As usual, if $\varphi$ is a formula and all of its free variables occur in the sequence $\bar{x}$ then this formula
may be denoted by $\varphi(\bar{x})$.
If $\mcA$ is a $\sigma$-structure, $\varphi(\bar{x})$ is first-order formula over $\sigma$ and $\bar{a} \in A^{|\bar{x}|}$,
then the notation $\mcA \models \varphi(\bar{a})$ has the same meaning as in first-order logic.

A {\em directed acyclic graph (DAG)}
is a pair $(V, E)$ where $V$ (the set of its vertices) is any set and $E \subseteq V \times V$ 
(the set of its edges) has the property that $(v, v) \notin E$
for all $v \in V$ and for all integers $n > 0$
there do not exist $v_0, \ldots, v_n \in V$ such that $(v_n, v_0) \in E$ and $(v_i, v_{i+1}) \in E$ for all $i< n$.
Suppose that $\mcG = (V, E)$ is a DAG.
If $v \in V$ then $\mr{par}(v)$ denotes the set of {\em parents of} $v$, that is, the set of all vertices $w \in V$ such 
that $(w, v) \in E$.
For $v \in V$ then we define the {\em maximal path rank} of $v$, 
denoted $\mr{mp}(v)$, to be the maximal integer $n > 0$
such that there is a directed path $v_0, \ldots, v_n \in V$ (meaning that $(v_i, v_{i+1}) \in E$ for all $i$) 
with $v_n = v$. 
We define the {\em maximal path rank} of $\mcG$, denoted $\mr{mp}(\mcG)$, as
$\mr{mp}(\mcG) = \max(\mr{mp}(v) : v \in V)$.

\section{Probability logic with aggregation functions}\label{Probability logic with aggregation functions}

\noindent
Let $\sigma$ be a finite relational signature.

\begin{defin}\label{definition of first-order formulas}{\rm
(i) Constructions of the form `$x = y$' and `$R(x_1, \dots, x_r)$', where $x, y$, $x_1, \ldots, x_r$ are variables, 
and $R \in \sigma$ has arity $r$, 
are called {\em atomic first-order formulas (over $\sigma$)}. 
By a {\em first-order literal (over $\sigma$)} we mean a first-order atomic formula (over $\sigma$) or a negation of such one.\\
(ii) If $\mcA$ is a $\sigma$-structure with domain $A$ and $a, b, a_1, \ldots, a_r \in A$, then the notation 
`$\mcA \models a = b$' and `$\mcA \models R(a_1, \ldots, a_r)$', where $R \in \sigma$,
have the same 
meaning as in first-order logic.
}\end{defin}

\begin{defin}\label{definition of atomic type} {\bf (Atomic $\sigma$-types)} {\rm
A consistent set $p$ of first-order literals over $\sigma$
is called an {\em atomic $\sigma$-type}.
If an atomic $\sigma$-type is denoted by $p(\bar{x})$ it is understood that every variable that occurs in a formula in $p(\bar{x})$
occurs in the sequence $\bar{x}$. 
An atomic $\sigma$-type $p(\bar{x})$ is called {\em complete} if for every first-order literal $\varphi(\bar{x}) \in PLA(\sigma)$,
either $\varphi(\bar{x})$ or $\neg\varphi(\bar{x})$ belongs to $p(\bar{x})$.
If $p(\bar{x})$ is an atomic $\sigma$-type and $\rng(\bar{y}) \subseteq \rng(\bar{x})$, then
$p(\bar{x}) \uhrc \bar{y}$ (or $p \uhrc \bar{y}$) denotes the set of all formulas $\varphi \in p(\bar{x})$
such that every variable of $\varphi$ occurs in $\bar{y}$.
}\end{defin}

\noindent
When convenient we will identify an atomic $\sigma$-type $p(\bar{x})$ with the formula obtained by taking the 
conjunction of all formulas in $p(\bar{x})$.
With this convention, if $\mcA$ is a $\sigma$-structure and $\bar{a} \in A^{|\bar{x}|}$ the notation
$\mcA \models p(\bar{a})$ makes sense and means, with model theoretic language, that $\bar{a}$ {\em realizes} $p(\bar{x})$ 
(in the structure $\mcA$).
Note that if $\sigma = \es$, then an atomic $\sigma$-type $p(\bar{x})$ will only contain literals of the form
$z = y$ or $z \neq y$ where $z, y \in \rng(\bar{x})$.

\begin{defin}\label{syntax of PLA}{\bf (Syntax of $PLA(\sigma)$)} {\rm 
By the {\em probability logic with aggregation functions over $\sigma$}, denoted $PLA(\sigma)$, we mean the set of objects called {\em formulas}
which are constructed as described below. 
We assume that we have an infinite set of symbols called variables, usually denoted $x, y, z, u, v$, possibly with indices.
For each $\varphi \in PLA(\sigma)$ the notation $Fv(\varphi)$ denotes the set of {\em free variables} of $\varphi$.
If we denote a formula by $\varphi(\bar{x})$, where $\bar{x}$ is a sequence of variables, it is understood that all free
variables of $\varphi(\bar{x})$ occur in $\bar{x}$.
\begin{enumerate}
\item  For each $c \in [0, 1]$, $c \in PLA(\sigma)$ (i.e. $c$ is a formula) and $c$ has no free variables. We may also let $\bot$ and $\top$
denote $0$ and $1$, respectively.

\item For all variables $x$ and $y$, `$x = y$' belongs to $PLA(\sigma)$. The free variables of `$x = y$'  are $x$ and $y$.

\item For every $R \in \sigma$, say of arity $r$, and any choice of variables $x_1, \ldots, x_r$, $R(x_1, \ldots, x_r)$ belongs to $PLA(\sigma)$.
The free variables of $R(x_1, \ldots, x_r)$ are $x_1, \ldots, x_r$.

\item If $\varphi, \psi, \chi \in PLA(\sigma)$ then the following also belong to $PLA(\sigma)$:
\begin{align*}
&(\neg\varphi), \ (\varphi \wedge \psi), \ (\varphi \vee \psi), \ (\varphi \to \psi), \text{ and (the {\em $\varphi$-weighted mean of $\psi$ and $\chi$})}\\
&(\varphi \psi + (1 - \varphi)\chi),
\end{align*}
but we may skip some parantheses if there is no ambiguity.
In each case the set of free variables of the new formula is the union of the sets of free variable of the formulas which it is constructed from.
We consider $\varphi \leftrightarrow \psi$
as an abbreviation of $(\varphi \to \psi) \wedge (\psi \to \varphi)$.

\item If $k \in \mbbN^+$, $\varphi_1(\bar{x}, \bar{y}), \ldots, \varphi_k(\bar{x}, \bar{y}) \in PLA(\sigma)$,
$p^=(\bar{x}, \bar{y})$ is a complete atomic $\es$-type (i.e. a complete description of the equalities and nonequalities between the variables),
where $\bar{x}$ and $\bar{y}$ are sequences of distinct variables such that $\rng(\bar{x}) \cap \rng(\bar{y}) = \es$
and $F : \big( [0, 1]^{<\omega} \big)^k \to [0, 1]$ is an aggregation function,
then 
\[
F(\varphi_1(\bar{x}, \bar{y}), \ldots, \varphi_k(\bar{x}, \bar{y}) : \bar{y} : p^=(\bar{x}, \bar{y}))
\]
belongs to $PLA(\sigma)$.
If this new formula is denoted $\psi$ then 
\[Fv(\psi) = \big( \bigcup_{i=1}^k Fv(\varphi_i)\big) \setminus \rng(\bar{y}),
\] 
thus this construction binds the variables in $\bar{y}$.
\end{enumerate}
}\end{defin}

\begin{defin}\label{definition of basic probability formula}{\rm
(i) A formula that does not contain any aggregation function is called {\em aggregation-free}.\\
(ii) If $n \in \mbbN^+$, $\alpha_1, \ldots, \alpha_n \in [0, 1]$ and $\psi_1(\bar{x}), \ldots, \psi_n(\bar{x}) \in PLA(\sigma)$ 
are such that each $\psi_i$ 
is a conjunction of first-order literals, then the formula $\bigwedge_{i=1}^n \big(\psi_i(\bar{x}) \rightarrow \alpha_i\big)$
is called a {\em basic probability formula}.
}\end{defin}

\begin{defin}\label{semantics of PLA}{\bf (Semantics of $PLA(\sigma)$)} {\rm
For each $\sigma$-structure $\mcA$, each formula $\varphi(\bar{x}) \in PLA(\sigma)$ and every $\bar{a} \in A^{|\bar{x}|}$,
we define a real number, denoted
$\mcA(\varphi(\bar{a}))$,  in the interval $[0, 1]$, called the 
{\em value of $\varphi(\bar{a})$ in $\mcA$}, as follows
(where if $\varphi$ has no free variable we just omit $\bar{x}$ and $\bar{a}$):

\begin{enumerate}
\item For every $c \in [0, 1]$ and every $\sigma$-structure $\mcA$, $\mcA(c) = c$.

\item For every $\sigma$-structure $\mcA$ and all $a, b \in A$, $\mcA(a = b) = 1$ if $\mcA \models a = b$ and otherwise $\mcA(a = b) = 0$.

\item For every $R \in \sigma$, of arity $r$ say, every $\sigma$-structure $\mcA$ and all $\bar{a} \in A^r$,
$\mcA(R(\bar{a})) = 1$ if $\mcA \models R(\bar{a})$ and otherwise $\mcA(R(\bar{a})) = 0$.

\item If $\varphi(\bar{x}), \psi(\bar{x}), \chi(\bar{x}) \in PLA(\sigma)$, $\mcA$ is a $\sigma$-structure and $\bar{a} \in A^{|\bar{x}|}$, then
\begin{align*}
&\mcA(\neg\varphi(\bar{a})) = 1 - \mcA(\varphi(\bar{a})), \\
&\mcA(\varphi(\bar{a}) \wedge \psi(\bar{a})) = \min(\mcA(\varphi(\bar{a})), \mcA(\psi(\bar{a}))), \\
&\mcA(\varphi(\bar{a}) \vee \psi(\bar{a})) = \max(\mcA(\varphi(\bar{a})), \mcA(\psi(\bar{a}))), \\
&\mcA(\varphi(\bar{a}) \to \psi(\bar{a})) = \min\big(1, \ 1 - \mcA(\varphi(\bar{a})) + \mcA(\psi(\bar{a}))\big), \text{ and} \\
&\mcA(\varphi(\bar{a})\psi(\bar{a}) + (1 - \varphi(\bar{a}))\chi(\bar{a})) = 
\mcA(\varphi(\bar{a}))\mcA(\psi(\bar{a})) + (1 - \mcA(\varphi(\bar{a})))\mcA(\chi(\bar{a})).
\end{align*}

\item If $k \in \mbbN^+$, $\bar{x}$ and $\bar{y}$ are sequences of distinct variables such that 
$\rng(\bar{x}) \cap \rng(\bar{y}) = \es$, 
$\varphi_1(\bar{x}, \bar{y}), \ldots, \varphi_k(\bar{x}, \bar{y}) \in PLA(\sigma)$,
$p^=(\bar{x}, \bar{y})$ is a complete atomic $\es$-type,
$F : \big([0, 1]^{<\omega}\big)^k \to [0, 1]$ is an aggregation function, 
$\mcA$ is a finite $\sigma$-structure and $\bar{a} \in A^{|\bar{x}|}$, then:
\[
\mcA\big(F(\varphi_1(\bar{a}, \bar{y}), \ldots, \varphi_k(\bar{a}, \bar{y}) : \bar{y} : p^=(\bar{a}, \bar{y}))\big) = 
F(\bar{r}_1, \ldots, \bar{r}_k)
\]
if there is some $\bar{b} \in A ^{|\bar{y}|}$ such that
$p^=(\bar{a}, \bar{b})$ holds
and, for $i = 1, \ldots, k$, 
\[
\bar{r}_i = \big(\mcA(\varphi_i(\bar{a}, \bar{b})) : \bar{b} \in A^{|\bar{y}|} \text{ and $p^=(\bar{a}, \bar{b})$ holds} \big),
\]
and otherwise 
$\mcA\big(F(\varphi_1(\bar{a}, \bar{y}), \ldots, \varphi_k(\bar{a}, \bar{y}) : \bar{y} : p^=(\bar{a}, \bar{y}))\big) = 0$.
\end{enumerate}
}\end{defin}

\noindent
With the above definitions the semantics of the propositional constructions in part~(4) coincide with
the common semantics for $\neg, \wedge, \vee$ and $\to$ when the values are 0 or 1.
Also, each propositional construction corresponds to a uniformly continuous function and this is essential in the proofs of the
main results.

\begin{rem}\label{remark on aggregations without identity constraints} 
{\bf (On aggregations without identity constraints)} {\rm
The reader may ask why we did not, for a $k$-ary aggregation function $F$, add to $PLA(\sigma)$ formulas of the form
\[
F(\varphi_1(\bar{x}, \bar{y}), \ldots, \varphi_k(\bar{x}, \bar{y}) : \bar{y})
\]
with the same semantics as in part~(5) of
Definition~\ref{semantics of PLA}
 except for omitting the condition that $p^=(\bar{a}, \bar{b})$ holds
(as $p^=$ no longer appears in the new formula).
The reason is that for formulas of this form our proof that an {\em admissible} aggregation function $F$
can be asymptotically eliminated
does not work out (where the notion `admissible' is defined in 
Definition~\ref{alternative definition of admissible function}), because the ``degrees of freedom'' which are determined
by an identity type (usually denoted $p^=$ here) matter in this context. 
More detailed explanations of why the proof does not work out are found in 
Remark~\ref{remark on necessity of p-equality}.
However, as we show in \cite{KW2}, every {\em strongly admissible} aggregation function $F$
(including the arithmetic and geometric means, but not max and min) can be asymptotically eliminated 
from a formula like $F(\varphi_1(\bar{x}, \bar{y}), \ldots, \varphi_k(\bar{x}, \bar{y}) : \bar{y})$
and for more general kinds of probability distributions than considered here.
}\end{rem}

\begin{defin}\label{definition of equivalent}{\rm
We say that $\varphi(\bar{x}) \in PLA(\sigma)$ and $\psi(\bar{x}) \in PLA(\sigma)$ are {\em equivalent} if,
for every finite $\sigma$-structure $\mcA$ and every $\bar{a} \in A^{|\bar{x}|}$, $\mcA(\varphi(\bar{a})) = \mcA(\psi(\bar{a}))$.
}\end{defin}

\begin{rem}\label{remark on basic probability sentences} {\rm
A basic probability formula which is also a sentence, that is, a formula without free variables, 
has the form $\bigwedge_{i=1}^n(\top \to c_i)$ where $c_i \in [0, 1]$ (and recall that $\top = 1$).
The formula $\bigwedge_{i=1}^n(\top \to c_i)$ is equivalent to $c$ where $c = \min\{c_1, \ldots, c_n\}$,
so every basic probability sentence is equivalent to a sentence of the form $c$ for some $c \in [0, 1]$.
}\end{rem}

\begin{defin}\label{definition of aggregation rank} {\rm
The {\em aggregation rank} of a formula $\varphi \in PLA(\sigma)$, denoted $\mr{agr}(\varphi)$, is defined as follows:
\begin{enumerate}
\item If $\varphi$ is aggregation-free then $\mr{agr}(\varphi) = 0$.
\item $\mr{agr}(\varphi \wedge \psi) = \mr{agr}(\varphi \vee \psi) = 
\mr{agr}(\varphi \rightarrow \psi) ) = \max\{\mr{agr}(\varphi), \mr{agr}(\psi)\}$.
\item $\mr{agr}\big( F(\varphi_1(\bar{x}, \bar{y}), \ldots, \varphi_k(\bar{x}, \bar{y}) : \bar{y} : p^=(\bar{x}, \bar{y})) \big) =
\max\{\mr{agr}(\varphi_i) : i = 1, \ldots, k \} + |\bar{y}|$.
\end{enumerate}
}\end{defin}

\begin{lem}\label{quantifier-free formulas are equivalent to bpf}
If $\varphi(\bar{x}) \in PLA(\sigma)$ is aggregation-free then $\varphi(\bar{x})$ is equivalent to a basic probability formula.
\end{lem}

\noindent
{\bf Proof.}
Let $p_1(\bar{x}), \ldots, p_m(\bar{x})$ enumerate, without repetition, all complete atomic $\sigma$-types in the variables $\bar{x}$.
As $\varphi(\bar{x})$ is aggregation-free it is clear that for all $i = 1, \ldots, m$, 
every $\sigma$-structure $\mcA$ and all $\bar{a}, \bar{b} \in A^{|\bar{x}|}$,
if $\mcA \models p_i(\bar{a}) \wedge p_i(\bar{b})$ then $\mcA(\varphi(\bar{a})) = \mcA(\varphi(\bar{b}))$.
Therefore there are (not necessarily distinct) $c_1, \ldots, c_m \in [0, 1]$ such that
whenever $\mcA$ is a $\sigma$-structure and $\bar{a} \in A^{|\bar{x}|}$, then, for every $i = 1, \ldots, m$,
\begin{align*}
&\text{if } \mcA \models p_i(\bar{a}) \text{ then } \mcA(\varphi(\bar{a})) = c_i, \text{ and} \\
&\mcA(p_i(\bar{a}) \rightarrow c_i) = 
\begin{cases}
c_i \text{ if } \mcA \models p_i(\bar{a}),\\
1 \text{ if } \mcA \not\models p_i(\bar{a}).
\end{cases}
\end{align*}
Thus, if $\mcA$ is a $\sigma$-structure and $\bar{a} \in A^{|\bar{x}|}$, then there is a unique $j$ such that
$\mcA \models p_j(\bar{a})$ and we get 
\[
\mcA\bigg(\bigwedge_{i=1}^m (p_i(\bar{a}) \rightarrow c_i)\bigg) \ = \ 
\min\big(\mcA(p_i(\bar{a}) \rightarrow c_i) : i = 1, \ldots, m\big) \ = \ c_j \ = \ 
\mcA(\varphi(\bar{a})).
\]
Hence $\varphi(\bar{x})$ is equivalent to the basic probability formula $\bigwedge_{i=1}^m (p_i(\bar{x}) \rightarrow c_i)$.
\hfill $\square$
\\

\noindent
Below we note that $PLA(\sigma)$ respects isomorphism.

\begin{lem}\label{isomorphisms preserve the value} {\bf (Truth value invariance under isomorphisms)}
Let $\mcA$ and $\mcB$ be isomorphic $\sigma$-structures and let $f$ denote an isomorphism from $\mcA$ to $\mcB$.
If $\varphi(\bar{x}) \in PLA(\sigma)$ and $\bar{a} \in A^{|\bar{x}|}$, then 
$\mcA(\varphi(\bar{a})) = \mcB(\varphi(f(\bar{a})))$.
\end{lem}

\noindent
{\bf Proof.}
We use induction on the complexity of $\varphi(\bar{x}) \in PLA(\sigma)$.
First suppose that $\varphi(\bar{x})$ is aggregation-free.
By Lemma~\ref{quantifier-free formulas are equivalent to bpf},
$\varphi(\bar{x})$ is equivalent to a basic probability formula and from the definition of a basic probability formula
it is immediate that $\mcA(\varphi(\bar{a})) = \mcB(\varphi(f(\bar{a})))$.
It is also clear that if $p^=(\bar{x}, \bar{y})$ is an atomic $\es$-type, then
$p^=(\bar{a}, \bar{b})$ holds if and only if $p^=(f(\bar{a}), f(\bar{b}))$ holds.

Now suppose that $\varphi(\bar{x})$ has the form
\[
F(\psi_1(\bar{x}, \bar{y}), \ldots, \psi_k(\bar{x}, \bar{y}) : \bar{y} : p^=(\bar{x}, \bar{y}))
\]
where $\rng(\bar{x}) \cap \rng(\bar{y}) = \es$.
If there is no $\bar{b} \in A^{|\bar{x}|}$ such that $p^=(\bar{a}, \bar{b})$ holds, then
no $\bar{b} \in B^{|\bar{x}|}$ exists such that $p^=(f(\bar{a}), \bar{b})$ holds,
and consequently $\mcA(\varphi(\bar{a})) = 0 = \mcB(\varphi(f(\bar{a})))$.

Now suppose that there is $\bar{b} \in A^{|\bar{x}|}$ such that $p^=(\bar{a}, \bar{b})$ holds.
By the induction hypothesis, we have, for all $i = 1, \ldots, k$, all $\bar{a} \in A^{|\bar{x}|}$ and all $\bar{b} \in A^{|\bar{y}|}$,
$\mcA(\psi_i(\bar{a}, \bar{b})) = \mcB(\psi_i(f(\bar{a}), f(\bar{b}))$.
For every $i$, let 
\begin{align*}
&\bar{r}_{i, \mcA} = (\mcA(\psi_i(\bar{a}, \bar{b})) : \bar{b} \in A^{|\bar{x}|} \text{ and $p^=(\bar{a}, \bar{b})$ holds}), \\
&\bar{r}_{i, \mcB} = (\mcB(\psi_i(f(\bar{a}), \bar{b})) : \bar{b} \in B^{|\bar{x}|} \text{ and $p^=(f(\bar{a}), \bar{b})$ holds}).
\end{align*}
Then, as $f$ is an isomorphism,
for every $c \in [0, 1]$, if $c$ occurs exactly $m$ times in $\bar{r}_{i, \mcA}$, then $c$ occurs exactly $m$ times in $\bar{r}_{i, \mcB}$.
Since $F$ is an aggregation function we get
$F(\bar{r}_{1, \mcA}, \ldots, \bar{r}_{k, \mcA}) = F(\bar{r}_{1, \mcB}, \ldots, \bar{r}_{k, \mcB})$ and hence
\begin{align*}
&\mcA\big(F(\psi_1(\bar{a}, \bar{y}), \ldots, \psi_k(\bar{a}, \bar{y}) : \bar{y} : p^=(\bar{a}, \bar{y}))\big) \ = \\ 
&\mcB\big(F(\psi_1(f(\bar{a}), \bar{y}), \ldots, \psi_k(f(\bar{a}), \bar{y}) : \bar{y} : p^=(\bar{a}, \bar{y}))\big).
\end{align*}
\hfill $\square$

\section{Directed parametrized probabilistic graphical models and induced sequences of probability distributions}\label{PGM}

\subsection{Sequences of probability distributions and asymptotic equivalence}

\noindent
Throughout this section (as in the rest of the article) we assume that $\sigma$ is a finite relational signature and that $\mbW_n$
denotes the set of all $\sigma$-structures with domain $[n]$.

\begin{defin}\label{definition of asymptotic probability distribution}{\rm
By a {\em sequence of probability distributions} (on $(\mbW_n : n \in \mbbN^+)$) we mean a sequence
$(\mbbP_n : n \in \mbbN^+)$ such that for every $n$, $\mbbP_n$ is a probability distribution on $\mbW_n$.
}\end{defin}

\begin{defin}\label{definition of asymptotically equivalent formulas}{\rm
Let $\varphi(\bar{x}), \psi(\bar{x}) \in  PLA(\sigma)$ where $\bar{x}$ is a tuple of distinct variables.
We say that $\varphi(\bar{x})$ and $\psi(\bar{x})$ are {\em asymptotically equivalent (with respect to $(\mbbP_n : n \in \mbbN^+)$)} 
if for all $\varepsilon > 0$
\[
\mbbP_n\Big(\big\{\mcA \in \mbW_n : \text{ there is $\bar{a} \in A^{|\bar{x}|}$ such that 
$|\mcA(\varphi(\bar{a})) - \mcA(\psi(\bar{a}))| > \varepsilon$}\big\} \Big) \to 0
\]
as $n \to \infty$.
}\end{defin}

\noindent
The following lemma is essential for the proof of the main results.

\begin{lem}\label{preservation of asymptotic equivalence under connectives} 
{\bf (Preservation of asymptotic equivalence under connectives)} 
Let $\varphi(\bar{x}), \varphi'(\bar{x}), \psi(\bar{x}), \psi'(\bar{x}), \chi(\bar{x}), \chi'(\bar{x}) \in PLA(\sigma)$.
Suppose that, with respect to $(\mbbP_n : n \in \mbbN^+)$, 
$\varphi(\bar{x})$ is asymptotically equivalent to $\varphi'(\bar{x})$,
$\psi(\bar{x})$ is asymptotically equivalent to $\psi'(\bar{x})$, and
$\chi(\bar{x})$ is asymptotically equivalent to $\chi'(\bar{x})$.
Let $\theta(\bar{x})$ be a formula constructed from $\varphi(\bar{x}), \psi(\bar{x})$ and/or $\chi(\bar{x})$
by any one of the constructions in part~(4) of 
Definition~\ref {syntax of PLA}
and let $\theta'(\bar{x})$ be constructed in the same way from $\varphi'(\bar{x}), \psi'(\bar{x})$ and/or $\chi'(\bar{x})$
(so for example $\theta(\bar{x})$ and $\theta'(\bar{x})$ could be $\varphi(\bar{x}) \wedge \psi(\bar{x}, respectively)$
and $\varphi'(\bar{x}) \wedge \psi'(\bar{x})$).
Then $\theta(\bar{x})$ and $\theta'(\bar{x})$ are asymptotically equivalent with respect to $(\mbbP_n : n \in \mbbN^+)$.
\end{lem}

\noindent
{\bf Proof.}
The functions $1 - x$, $\max(x, y)$, $\min(x, y)$, $\min(1, 1 - x + y)$, and $x\cdot y + (1-x)\cdot z$
are uniformly continuous. 
Therefore the conclusion follows from the semantics (Definition~\ref{semantics of PLA}),
the assumptions about asymptotic equivalence and from the assumption that
$\theta$ and $\theta'$ are constructed in the same way from $\varphi(\bar{x}), \psi(\bar{x})$ and/or $\chi(\bar{x})$
and from $\varphi'(\bar{x}), \psi'(\bar{x})$ and/or $\chi'(\bar{x})$, respectively.
\hfill $\square$

\subsection{Conditional probability logic and lifted Bayesian networks}

\noindent
The para\-metri\-zed probabilistic graphical model that we will use was introduced in
\cite{Kop20} and is called {\em lifted Bayesian network}. 
Lifted Bayesian networks use, in their definition, a logic (also introduced in \cite{Kop20})
called {\em conditional probability logic ($CPL$)} so we introduce this logic first.
(There are previously considered logics, such as the ones in \cite{Bac} and \cite{Hal},  
the expressivity of which is at least as strong as the expressivity of $CPL$,
but for such logics we have not found any ``convergence'' results of the kind proved in \cite{Kop20} which will be used later.)

\begin{defin}\label{definition of CPL} {\bf (Syntax of $CPL(\sigma)$)} {\rm
Let $\sigma$ be a signature.
The set of {\em conditional probability formulas over $\sigma$}, denoted $CPL(\sigma)$, 
is defined as follows:
\begin{enumerate}
\item Every atomic $\sigma$-formula belongs to $CPL(\sigma)$ (where `atomic' has the same meaning as in first-order logic with equality).

\item If $\varphi, \psi \in CPL(\sigma)$ then $(\neg\varphi), (\varphi\wedge\psi), (\varphi\vee\psi), (\varphi\rightarrow\psi), 
(\varphi\leftrightarrow\psi), (\exists x \varphi) \in CPL(\sigma)$ where $x$ is a variable.

\item If $r \geq 0$ is a real number, $\varphi, \psi, \theta, \tau \in CPL(\sigma)$ and $\bar{y}$ is a sequence of distinct variables, then
\begin{align*}
&\Big( r + \| \varphi \ |  \ \psi \|_{\bar{y}} \ \geq \ 
\| \theta \ |  \ \tau \|_{\bar{y}} \Big) \in CPL(\sigma)
\ \ \text{ and} \\
&\Big( \| \varphi \ |  \ \psi \|_{\bar{y}} \ \geq \ 
\| \theta \ |  \ \tau \|_{\bar{y}} + r \Big) \in CPL(\sigma).
\end{align*}
In both these new formulas all variables of $\varphi, \psi, \theta$ and $\tau$ that appear in the sequence 
$\bar{y}$ become {\em bound}. That this construct is a form of quantification becomes apparent from
its semantics below.
\end{enumerate}
}\end{defin}

\noindent
A formula $\varphi \in CPL(\sigma)$ is called {\em quantifier-free} if it is constructed from atomic formulas by
using only connectives $\neg, \wedge, \vee, \rightarrow$ and $\leftrightarrow$.

\begin{defin}\label{semantics of CPL} {\bf (Semantics of $CPL(\sigma)$)} {\rm
\begin{enumerate}
\item The interpretations of $\neg, \wedge, \vee, \rightarrow, \leftrightarrow$ and $\exists$ are as in first-order logic.

\item Suppose that $\mcA$ is a {\em finite} $\sigma$-structure and let $\varphi(\bar{x}, \bar{y}), \psi(\bar{x}, \bar{y}), 
\theta(\bar{x}, \bar{y}), \tau(\bar{x}, \bar{y}) \in CPL(\sigma)$.
Let $\bar{a} \in A^{|\bar{x}|}$.
\begin{enumerate}
\item We define $\varphi(\bar{a}, \mcA) = \big\{\bar{b} \in A^{|\bar{y}|} : \mcA \models \varphi(\bar{a}, \bar{b}) \big\}$.

\item The expression 
\[
\mcA \  \models \ 
\Big( r + \| \varphi(\bar{a}, \bar{y}) \ | \ \psi(\bar{a}, \bar{y}) \|_{\bar{y}} \ \geq \ 
\| \theta(\bar{a}, \bar{y}) \ | \ \tau(\bar{a}, \bar{y}) \|_{\bar{y}} \Big)
\]
means that $\psi(\bar{a}, \mcA) \neq \es$, $\tau(\bar{a}, \mcA) \neq \es$ and
\[
r + \frac{\big| \varphi(\bar{a}, \mcA) \cap \psi(\bar{a}, \mcA) \big|}{\big| \psi(\bar{a}, \mcA) \big|} \ \geq \ 
\frac{\big| \theta(\bar{a}, \mcA) \cap \tau(\bar{a}, \mcA) \big|}{\big| \tau(\bar{a}, \mcA) \big|}
\]
and in this case we say that 
$\Big( r + \| \varphi(\bar{a}, \bar{y}) \ | \ \psi(\bar{a}, \bar{y}) \|_{\bar{y}} \ \geq \ 
\| \theta(\bar{a}, \bar{y}) \ | \ \tau(\bar{a}, \bar{y}) \|_{\bar{y}} \Big)$
is true (or holds) in $\mcA$.
If $\psi(\bar{a}, \mcA) = \es$ or $\tau(\bar{a}, \mcA) = \es$ or
\[
r + \frac{\big| \varphi(\bar{a}, \mcA) \cap \psi(\bar{a}, \mcA) \big|}{\big| \psi(\bar{a}, \mcA) \big|} \ < \ 
\frac{\big| \theta(\bar{a}, \mcA) \cap \tau(\bar{a}, \mcA) \big|}{\big| \tau(\bar{a}, \mcA) \big|}
\]
then we write
\[
\mcA \  \not\models \ 
\Big( r + \| \varphi(\bar{a}, \bar{y}) \ | \ \psi(\bar{a}, \bar{y}) \|_{\bar{y}} \ \geq \ 
 \| \theta(\bar{a}, \bar{y}) \ | \ \tau(\bar{a}, \bar{y}) \|_{\bar{y}} \Big)
\]
and say that 
$\Big( r + \| \varphi(\bar{a}, \bar{y}) \ | \ \psi(\bar{a}, \bar{y}) \|_{\bar{y}} \ \geq \ 
\| \theta(\bar{a}, \bar{y}) \ | \ \tau(\bar{a}, \bar{y}) \|_{\bar{y}} \Big)$
is false in $\mcA$.

\item The meaning of 
\[
\mcA \  \models \ 
\Big( \| \varphi(\bar{a}, \bar{y}) \ | \ \psi(\bar{a}, \bar{y}) \|_{\bar{y}} \ \geq \ 
\| \theta(\bar{a}, \bar{y}) \ | \ \tau(\bar{a}, \bar{y}) \|_{\bar{y}} + r \Big)
\]
is defined similarly.
\end{enumerate}
\end{enumerate}
}\end{defin}

\noindent

\begin{defin}\label{definition of BN} {\bf (Lifted Bayesian network)} {\rm
Let $\sigma$ be a finite relational signature.
A {\em lifted Bayesian network for $\sigma$} is determined by the following components:
\begin{itemize}
\item[(a)] An acyclic directed graph (DAG) $\mbbG$ with vertex set $\sigma$.

\item[(b)] For each $R \in \sigma$, a number $\nu_R \in \mbbN^+$, formulas $\chi_{R, i}(\bar{x}) \in CPL(\mr{par}(R))$,
for $i = 1, \ldots, \nu_R$, where $|\bar{x}|$ equals the arity of $R$, such that
$\forall \bar{x} \big( \bigvee_{i = 1}^{\nu_R} \chi_{R, i}(\bar{x})\big)$ is valid (i.e. true in all $\mr{par}(R)$-structures) and if
$i \neq j$ then $\exists \bar{x} \big(\chi_{R, i}(\bar{x}) \wedge \chi_{R, j}(\bar{x})\big)$
is unsatisfiable.
Each $\chi_{R, i}$ will be called an {\em aggregation formula (of $\mbbG$)}.

\item[(c)] For each $R \in \sigma$ and each $1 \leq i \leq \nu_R$, 
a number denoted $\mu(R \ | \ \chi_{R, i})$ (or $\mu(R(\bar{x}) \ | \ \chi_{R, i}(\bar{x}))$)
in the interval $[0, 1]$.
\end{itemize}
}\end{defin}

\noindent
{\em We use the convention to denote a lifted Bayesian network by the same symbol (e.g. $\mbbG$) as its underlying DAG.}
Observe that Definition~\ref{definition of BN} makes sense if $\sigma$ is empty. In this case
the underlying DAG has empty vertex set (and edge set) and no numbers or formulas as
in parts~(b) and~(c) of the definition need to be specified.

\begin{defin}\label{the probability distribution induced by an LBN}
{\bf (The probability distribution induced by a lifted Bayesian network)} {\rm
Let $\sigma$ be a finite nonempty relational signature and let $\mbbG$ denote a 
lifted Bayesian network over $\sigma$.
In this definition we denote the arity of $R \in \sigma$ by $k_R$.
Suppose that the underlying DAG of $\mbbG$ has maximal path rank $\rho$.
Let $\sigma_{-1} = \es$ and, for $0 \leq r \leq \rho$, let $\sigma_r = \{R \in \sigma : \mr{mp}(R) \leq r\}$.
For $r = -1, 0, 1, \ldots, \rho$, let $\mbbG_r$ be the subnetwork of $\mbbG$ which is
induced by $\sigma_r$ and let $\mbW^r_n$ be the set of all $\sigma_r$-structures 
domain $[n]$.
Note that $\mbbG_\rho = \mbbG$ and $\mbW^\rho_n = \mbW_n$.
Let $\mbbP^{-1}_n$ be the unique probability distribution on the singleton set $\mbW^{-1}_n$.

By induction on $r$ we define, for every $r = 0, 1, \ldots, \rho$, a probability distribution $\mbbP^r_n$ on the set $\mbW^r_n$ as follows:
For every $\mcA \in \mbW^r_n$,
\[
\mbbP^r_n(\mcA) \ = \ \mbbP^{r-1}_n(\mcA \uhrc \sigma_{r-1}) 
\prod_{R \in \sigma_r \setminus \sigma_{r-1}} \ \prod_{i=1}^{\nu_R} \ \prod_{\bar{a} \in \chi_{R, i}(\mcA \uhr \sigma_{r-1})} 
\lambda(\mcA, R, i, \bar{a})
\]
where
\[
\lambda(\mcA, R, i, \bar{a}) = 
\begin{cases}
\mu(R \ | \ \chi_{R, i}) \ \ \ \ \ \ \ \text{ if } \mcA \models \chi_{R, i}(\bar{a}) \wedge R(\bar{a}),\\
1 - \mu(R \ | \ \chi_{R, i}) \ \ \text{ if } \mcA \models \chi_{R, i}(\bar{a}) \wedge \neg R(\bar{a}),\\
\text{0 \ \ \ \ \ \ \ \ \ \ \ \ \ \ \ \ \ \ \ \ otherwise.}
\end{cases}
\]
Finally we let $\mbbP_n = \mbbP^\rho_n$ so 
$(\mbbP_n : n \in \mbbN^+)$ is a sequence of probability distributions on $(\mbW_n : n \in \mbbN^+)$ which we call
the {\em sequence of probability distributions induced by $\mbbG$}.
}\end{defin}

\section{Expressivity}\label{Discussion}

\noindent
The scope of the results shown here depends both on the expressivity of the query language $PLA$ and also on the expressivity of the lifted Bayesian networks which induce the probability distributions for which we show our results.

\subsection{Scope of the underlying families of probability distributions}\label{ExpressLBN}
Conditional probability logic allows the expression of discrete conditions based on relative frequencies, in addition to the full power of first-order logic.
Such conditions are often used as triggers in policy or engineering applications.
For instance, consider modelling infectious disease dynamics on networks. Then $CPL$ can express a variety of trigger conditions, such as the occurrence of a single positive case (using existential quantification) or a certain percentage of people being infected (using relative frequency quantification).
Lifted Bayesian networks then allow the modelling of actions that may be taken when those conditions are met. This suffices for modelling a variety of real-world policy decisions (such as those summarised in \cite[Table II]{BissetCDFMM14}).  For further examples of the expressivity of $CPL$ see
Example~3.5 and remarks~3.4 and~3.6 in~\cite{Kop20}

A clearly important fragment of $CPL$ for which our results hold is first-order logic itself. Lifted Bayesian networks whose formulas are first-order already suffice to model the relational Bayesian network specifications of Cozman and Maua's probabilistic finite model theory \cite{CM18,CM19}.

Beyond $CPL$ and lifted Bayesian networks, our results generalize immediately to every other 
sequence of probability distributions that is asymptotically equivalent to a sequence of distributions
induced by a lifted Bayesian network in 
the following sense:

\begin{defin}{\rm
Two sequences of distributions $\mbbP = (\mbbP_n : \in \mbbN^+)$ and 
$\mbbP' = (\mbbP'_n : \mbbN^+)$ are \emph{asymptotically equivalent} if 
$\stackrel[n\rightarrow\infty]{}{\lim}\underset{A\subseteq\mbW_n}{\sup}|\mbbP_n(A)-\mbbP'_n(A)|=0$.
}\end{defin}

\begin{rem}{\rm
In measure theoretic terms, the sequences of distributions $\mbbP$
and $\mbbP'$ are asymptotically equivalent if and only if the
limit of the total variation difference between them is $0$. 
}\end{rem}

For two formalisms that are very different from lifted Bayesian networks, 
namely probabilistic logic programming under the distribution semantics and functional lifted Bayesian networks,
it has recently been demonstrated~\cite{Wei21b, Wei21c} that every sequence of distributions induced by such a formalism 
is asymptotically equivalent to a sequence of distributions that is induced by a lifted 
Bayesian network in which all aggregation formulas are Boolean combinations of atomic formulas.

Probabilistic logic programming is one of the most-studied formalisms for statistical relational artificial intelligence that is unique in supporting recursion in the context of negation-by-failure, a feature inherited from classical logic programming. It has found significant practical application in bioinformatics~\cite{DK}.

Functional lifted Bayesian networks are much closer to $PLA$ itself, as they are designed to support continuous dependencies on relative frequency. However, unlike $PLA$ the aggregation functions used in defining a functional lifted Bayesian network must not be nested, which limits their expressivity but ensures the asymptotic equivalence to a quantifier-free lifted Bayesian network. They can model both linear and logistic regression functions, which suffices to express domain-size aware relational logistic regression~\cite{Wei21a}.

\begin{exam}\label{example of LBN} {\rm
To give some feeling of what kind of distributions can be described with lifted Bayesian networks, we consider the following
example which we describe informally.
Suppose we have properties $P_1, \ldots, P_s$ which also correspond to unary relation symbols. Each $P_i$ may be (conditionally) dependent
of some $P_j$ and (conditionally) independent of other $P_k$. 
These (conditional) dependencies and independencies can be described by a directed acyclic graph with vertex set $\{P_1, \ldots, P_s\}$.
To each $P_i$ we associate some $CPL$-formulas that use only $P_j$ among the parents of $P_i$ and which define cases such that
within each case $P_i(x)$ holds with a fixed probability.
Let $E$ be a binary relation symbol (corresponding to some relation) and let 
the probability that $E(x, y)$ holds depend (only) on which $P_i$ are satisfied by $x$ and $y$, respectively.
More formally the directed acyclic graph is enlarged with the vertex $E$ and arrows from $P_i$ to $E$ for all $P_i$
which have influence on the probability of $E$.
Let $R$ be a binary relation symbol,
let $0 < r_1 < \ldots < r_t = 1$ and $c_1, \ldots, c_t \in [0, 1]$.
Let the probability that $R(x, y)$ holds be $c_i$ if $\min(0, d_x - d_y) \in [r_i, r_{i+1})$ where 
$d_x$ is the proportion of $z$ with $E(z, x)$, among $z$ such that $E(u, z)$ for at least $1/10$ of the $u$ in the domain,
and $d_y$ is the proportion of $z$ with $E(z, y)$, among $z$ such that $E(u, z)$ for at least $1/10$ of the $u$ in the domain.
More formally, the directed acyclic graph is enlarged with a vertex $R$ and an arrow from $E$ to $R$, and
for every $i = 1, \ldots, t-1$, a $CPL$-formula which expresses that $\min(0, d_x - d_y) \in [r_i, r_{i+1})$ is associated to $R$.
}\end{exam}

\subsection{Expressivity of PLA}
$CPL$ is fundamentally distinct from $PLA$ by working in a 0--1--valued rather than a continuous-valued logic. $CPL$ therefore supports nesting conditional probability quantifiers, but does not allow for continuous dependencies on those conditional probabilities nor for other aggregation functions than conditional probabilities. 

The expressiveness of $PLA$ arises precisely from allowing nested combinations of different aggregation functions. The support for arithmetic mean, and variations of it, among them opens up new possibilities not covered by any of the aforementioned formalisms.

\begin{exam}\label{one stage example of rank}
{\bf (A measure of similarity)} {\rm
Let $E$ be a binary relation symbol. 
A measure of the {\em similarity of two elements $x$ and $y$} is given by considering the fraction of elements which have the same connections to $x$ and $y$. This can be expressed in $PLA$ by:
$\psi(x, y) := $
\begin{align*}
  \mathrm{am}( ( E(z,x) \leftrightarrow E(z,y) ) \wedge ( E(x,z) \leftrightarrow E(y,z) ): 
 z : x \neq y \wedge y \neq z \wedge x \neq z ).
\end{align*}
  ``The similarity to $x$ of the most similar other element'' is given by
\[
  \mathrm{max}( \psi(x,y): y : x \neq y ).
\]
  ``The average similarity of $x$ to other elements'' is given by
\[
  \mathrm{am}( \psi(x,y): y : x \neq y  ).
\]
  ``The lowest similarity score of any two elements'' is expressed by
\[
  \mathrm{min}(( \mathrm{min}( \psi(x,y): y : x \neq y )) : x : x = x ).
\]
In Example~\ref{SimRank} we show that all the stages of SimRank \cite{JW}
are expressible in $PLA$.
}\end{exam}

\begin{exam}\label{simple case of conditional arithmetic mean}
{\bf (Conditional arithmetic mean)} {\rm
There are situations when we are interested in the mean
over elements that satisfy some condition. 
In the present context we can express this situation by considering $PLA$-formulas
$\varphi(\bar{x}, \bar{y})$ and $\psi(\bar{x}, \bar{y})$ where $\psi$ is 0--1 valued. 
Let $p^=(\bar{x}, \bar{y})$ be a complete atomic $\es$-type, so it expresses all identity relations
among the variables $\bar{x}\bar{y}$.
For a finite structure $\mcA$ and $\bar{a} \in A^{|\bar{x}|}$, the arithmetic mean of 
$\mcA(\varphi(\bar{a}, \bar{b}))$ as $\bar{b}$ ranges over all tuples in $A^{|\bar{y}|}$ that
satisfy $p^=(\bar{a}, \bar{y})$ and $\psi(\bar{a}, \bar{y})$ can, letting
$X = \{\bar{b} \in A^{|\bar{y}|} : p^=(\bar{a}, \bar{b}) \text{ holds}\}$, be written as
\begin{equation}\label{conditional average in simple case}
\frac{\sum_{\bar{b} \in X} \mcA\big(\varphi(\bar{a}, \bar{b}) \wedge \psi(\bar{a}, \bar{b})\big)}
{\sum_{\bar{b} \in X} \mcA\big(\psi(\bar{a}, \bar{b})\big)}
\end{equation}
if at least one $\bar{b}$ satisfies $p^=(\bar{a}, \bar{y})$ and $\psi(\bar{a}, \bar{y})$.

We wish to find a $PLA$-formula $\theta(\bar{x})$ such that $\mcA(\theta(\bar{a}))$ 
equals~(\ref{conditional average in simple case}) whenever the denominator is positive.
For this we use the aggregation function `cam' defined for all $\bar{p}, \bar{q} \in [0, 1]^{<\omega}$
as follows, where $\sqrt{\bar{q}} = (\sqrt{q_1}, \ldots, \sqrt{q_n})$ if $\bar{q} = (q_1, \ldots, q_n)$:
\begin{align*}
&\text{$\mr{cam}(\bar{p}, \bar{q}) = 0$ if  $\bar{p}$ contains only zeros, and otherwise} \\
&\mr{cam}(\bar{p}, \bar{q})  =  \frac{\mr{am}(\bar{p})}{
\max(\mr{am}(\sqrt{\bar{p}}), \mr{am}(\bar{q}))}.
\end{align*}
Note that if $\bar{p}$ contains at least one nonzero entry, then $0 < \mr{am}(\bar{p}) \leq \mr{am}(\sqrt{\bar{p}})$
and hence $\mr{cam}(\bar{p}, \bar{q}) \leq \frac{\mr{am}(\bar{p})}{\mr{am}(\sqrt{\bar{p}})} \in (0, 1]$.
So the division with $\max(\mr{am}(\sqrt{\bar{p}}), \mr{am}(\bar{q}))$ instead of just $\mr{am}(\bar{q})$
makes sure that $\mr{cam}(\bar{p}, \bar{q})$ always belongs to $[0, 1]$, but also,
by Proposition~\ref{cam is admissible} below, it follows that
cam is {\em admissible} (that is, it has some ``continuity properties'') 
so that the main results apply to formulas using it.

Let $\theta(\bar{x})$ be the $PLA$-formula
\[
\mr{cam}\big(\varphi(\bar{x}, \bar{y}) \wedge \psi(\bar{x}, \bar{y}), \ 
\psi(\bar{x}, \bar{y}) : \bar{y} : p^=(\bar{x}, \bar{y})\big).
\]
Let 
\begin{align*}
&\bar{p} = \big( \mcA\big( \varphi(\bar{a}, \bar{b}) \wedge \psi(\bar{a}, \bar{b})\big)  : 
\bar{b} \in A^{|\bar{y}|} \text{ and $p^=(\bar{a}, \bar{b})$ holds}  \big), \\
&\bar{q} = \big( \mcA\big( \psi(\bar{a}, \bar{b})\big)  : 
\bar{b} \in A^{|\bar{y}|} \text{ and $p^=(\bar{a}, \bar{b})$ holds}  \big)
\end{align*}
and suppose that $\bar{q}$ is not constantly zero, so it contains at least one 1.
Since $\mcA\big( \varphi(\bar{a}, \bar{b}) \wedge \psi(\bar{a}, \bar{b})\big) \leq 
\mcA\big( \psi(\bar{a}, \bar{b})\big)$ for all $\bar{b}$, it follows that 
$\mr{am}(\sqrt{\bar{p}}) \leq \mr{am}(\bar{q})$ and hence
\[
\mcA(\theta(\bar{a})) = \mr{cam}(\bar{p}, \bar{q}) = \frac{\mr{am}(\bar{p})}{\mr{am}(\bar{q})}
\]
which equals~(\ref{conditional average in simple case}) under the stated assumptions.
}\end{exam}

\begin{exam}\label{conditional arithmetic mean} 
{\bf (Conditional arithmetic mean with relaxed identity constraints)} {\rm
Let $\varphi(\bar{x}, y, z), \psi(\bar{x}, y, z) \in PLA(\sigma)$ and assume that
$\psi$ is 0--1 valued.
Let $\mcA \in \mbW_n$ and  $\bar{a} \in [n]^{|\bar{x}|}$.
Suppose that we want to express the average of $\mcA(\varphi(\bar{a}, b, c))$
as $(b, c)$ ranges over all ordered pairs of elements in $[n] \setminus \rng(\bar{a})$
such that $\mcA(\psi(\bar{a}, b, c)) = 1$. In other words we allow that $b \neq c$ and that $b = c$
so we have not fixed a complete identity constraint on $y$ and $z$.
Then we cannot directly apply the methods of 
Example~\ref{simple case of conditional arithmetic mean}
since those methods require that we consider the conditional arithmetic mean only for
 $(b, c)$ such that $b\neq c$ or only for $(b, c)$ such that $b = c$.
However we can use the idea of 
Example~\ref{simple case of conditional arithmetic mean}
together with some additional ``tricks'' which we now explain.

If we let $X = [n] \setminus \rng(\bar{a})$, $n' = n - |\rng(\bar{a})|$ and
$\varphi'(\bar{x}, y, z)$ denotes $\varphi(\bar{x}, y, z) \wedge \psi(\bar{x}, y, z)$
then the described conditional average can be written as
\begin{align}\label{mean conditioned on psi}
&\frac{\sum_{(b, c) \in X^2} \mcA\big(\varphi'(\bar{a}, b, c)\big)}
{\sum_{(b, c) \in X^2} \mcA\big(\psi(\bar{a}, b, c)\big) }  \\ 
= \ 
&\frac{\frac{1}{n'(n'-1)}\sum_{\substack{(b, c) \in X^2 \\ b \neq c}} 
\mcA\big(\varphi'(\bar{a}, b, c)\big) \ + \ 
\frac{1}{n'(n'-1)}\sum_{b \in X} \mcA\big(\varphi'(\bar{a}, b, b)\big)}
{\frac{1}{n'(n'-1)}\sum_{\substack{(b, c) \in X^2 \\ b \neq c}} 
\mcA\big(\psi(\bar{a}, b, c)\big) \ + \ 
\frac{1}{n'(n'-1)}\sum_{b \in X} \mcA\big(\psi(\bar{a}, b, b)\big)} \nonumber \\
= \
&\frac{\mr{am}\big(\mcA\big(\varphi'(\bar{a}, b ,c)\big) : (b, c) \in X^2, b \neq c\big)
+ \frac{1}{n'}\mr{am}\big(\varphi'(\bar{a}, b ,b)\big) : (b, c) \in X^2, b \neq c\big)}
{\mr{am}\big(\mcA\big(\psi(\bar{a}, b ,c)\big) : (b, c) \in X^2, b \neq c\big)
+ \frac{1}{n'}\mr{am}\big(\psi(\bar{a}, b ,b)\big) : (b, c) \in X^2, b \neq c\big)}.  \nonumber
\end{align}
We wish to express the above by a $PLA$-formula that uses only admissible
aggregation functions, but a problem is that the expressions above
are undefined if $\psi(\bar{a}, b, c) = 0$ for all $(b, c) \in X^2$.

For every $\bar{r} = (r_1, \ldots, r_m) \in [0, 1]^{<\omega}$,
let $\sqrt{\bar{r}} = (\sqrt{r_1}, \ldots, \sqrt{r_m})$.
For $\bar{p}, \bar{q} \in [0, 1]^{<\omega}$ let $\bar{p}\bar{q}$ denote the {\em concatenation}
of $\bar{p}$ and $\bar{q}$.
Let $f^{-1}$ denote the inverse of the function $f(x) = x(x-1)$ on the positive reals.

For $\bar{p}, \bar{q}, \bar{r}, \bar{s} \in [0, 1]^{<\omega}$ define
\[
\text{$\mr{cam}^*(\bar{p}, \bar{q}, \bar{r}, \bar{s}) = 0$ if $\bar{p}$ and $\bar{q}$ contain only zeros,}
\]
and otherwise, letting $m = f^{-1}(|\bar{p}|)$ and
$\lambda(\bar{p}, \bar{r}) = 1 - \max(\bar{p}\bar{r})(1 - \frac{1}{m})$,
\begin{align}
&\mr{cam}^*(\bar{p}, \bar{q}, \bar{r}, \bar{s}) = \nonumber \\
&\frac{\max(\bar{p}\bar{r}) \mr{am}(\bar{p}) + \lambda(\bar{p}, \bar{r})\mr{am}(\bar{q})}
{\max\big[ \max(\bar{p}\bar{r}) \mr{am}(\sqrt{\bar{p}}) + 
\lambda(\bar{p}, \bar{r})\mr{am}(\sqrt{\bar{q}}), \ 
\max(\bar{p}\bar{r})\mr{am}(\bar{r}) + 
\lambda(\bar{p}, \bar{r})\mr{am}(\bar{s})\big] }. \label{definition of cam*, first occurrence}
\end{align}
Note that $\mr{cam}^*$ is defined for all possible 
$\bar{p}, \bar{q}, \bar{r}, \bar{s} \in [0, 1]^{<\omega}$ and that its output is always in $[0, 1]$.
By Proposition~\ref{cam is admissible} below,
$\mr{cam}^*$ is {\em admissible} and hence the main results apply to formulas using it.

Let $\theta(\bar{x})$ be the formula
\[
\mr{cam}^*(\varphi'(\bar{x}, y, z), \varphi'(\bar{x}, y, y), \psi(\bar{x}, y, z), \psi(\bar{x}, y, y) : y, z : 
p^=(\bar{x}, y, z))
\]
where $p^=(\bar{x}, y, z)$ expresses the identity relations among the elements in $\bar{a}$,
that $y$ and $z$ are different from all variables in $\bar{x}$, and that
$y \neq z$.

We claim that $\mcA(\theta(\bar{a}))$ equals~(\ref{mean conditioned on psi}) whenever
(\ref{mean conditioned on psi}) is well defined.
Let 
\begin{align*}
&\bar{p} = \big( \mcA(\varphi'(\bar{a}, b, c)) : (b, c) \in X^2, b \neq c \big), \\
&\bar{q} = \big( \mcA(\varphi'(\bar{a}, b, b)) : (b, c) \in X^2, b \neq c \big), \\
&\bar{r} = \big( \mcA(\psi(\bar{a}, b, c)) : (b, c) \in X^2, b \neq c \big), \\
&\bar{s} = \big( \mcA(\psi(\bar{a}, b, b)) : (b, c) \in X^2, b \neq c \big)
\end{align*}
and note that the length of each of the above sequences is $n'(n'-1)$.
Then $\mcA(\theta(\bar{a})) = cam^*(\bar{p}, \bar{q}, \bar{r}, \bar{s})$.
Suppose that at least one of $\bar{p}$ and $\bar{q}$ contain at least one non-zero entry.
Observe that $\mcA(\varphi'(\bar{a}, b, c)) \leq \mcA(\psi(\bar{a}, b, c))$ for all $b$ and $c$
and recall that $\mcA(\psi(\bar{a}, b, c))$ is either 0 or 1.

It follows that if some entry of $\bar{r}$ is 1 then $\max(\bar{p}\bar{r}) = 1$ and 
$\lambda(\bar{p}, \bar{r}) = 1/m$, so
\[
\mr{cam}^*(\bar{p}, \bar{q}, \bar{r}, \bar{s}) \ = \ 
\frac{\mr{am}(\bar{p}) + \frac{1}{m}\mr{am}(\bar{q})}
{ \mr{am}(\bar{r}) + \frac{1}{m}\mr{am}(\bar{s}) }
\]
which is equal to~(\ref{mean conditioned on psi}).

Now suppose that all entries of $\bar{r}$ are zero (so all entries of $\bar{p}$ are zero as well)
but some entry of $\bar{s}$ is one.
Then $\max(\bar{p}\bar{r}) = 0$ and $\lambda(\bar{p}, \bar{r}) = 1$ and hence
\[
\mr{cam}^*(\bar{p}, \bar{q}, \bar{r}, \bar{s}) \ = \ 
\frac{\mr{am}(\bar{q})}{\mr{am}(\bar{s})} \ = \ 
\frac{\sum_{b\in X} \mcA\big(\varphi'(\bar{a}, b, b)\big)}
{\sum_{b\in X} \mcA\big(\psi(\bar{a}, b, b)\big)}
\]
which also equals~(\ref{mean conditioned on psi}) in this case.

In the next example we show that $\mr{cam}^*$ can be used to define the ``stages'' of
the so-called SimRank.
}\end{exam}

\begin{exam}\label{SimRank} {\bf (SimRank)} {\rm
Consider a signature $\sigma$ with a binary relation symbol $E$. 
For a finite $\sigma$-structure $\mcA$,
a measure of the similarity of two elements $a, b \in A$ is given by the so-called
SimRank \cite{JW} defined recursively as
\[
s(a, b) = \frac{C \cdot \sum_{u \in I(a)} \sum_{v \in I(b)} s(u, v)}{|I(a)||I(b)|}
\]
where $C \in (0, 1]$ is a constant and $I(a)$ denotes the set of in-neighbours of $a$, that is,
$I(a) = \{c \in A : \mcA \models E(c, a)\}$.
The SimRank $s(a, b)$ can be estimated in stages by {\em $k$'th stage SimRanks}
$s_k(a, b)$
defined as follows
\[
s_0(a, b) = 
\begin{cases}
1 \ \ \text{ if } a = b \\
0 \ \ \text{ if } a \neq b,
\end{cases}
\]
and $s_{k+1}(a, b) = 1$ if $a = b$ and otherwise
\[
s_{k+1}(a, b) = \frac{C \cdot \sum_{u \in I(a)} \sum_{v \in I(b)} s_k(u, v)}{|I(a)||I(b)|}.
\]
Then $\lim_{k\to\infty} s_k(a, b) = s(a, b)$ \cite{JW}.

We now construct, for any $k \in \mbbN$, 
a $PLA(\sigma)$-formula $\varphi_k(x, y)$ such that for every finite $\sigma$-structure
$\mcA$ and all $a, b \in A$, $\mcA(\varphi_k(a, b)) = s_k(a, b)$.
For simplicity we let $C = 1$, because if we have defined $\varphi_k$ so that the above
holds for $C = 1$ then we can use the weighted mean 
(from Definition~\ref{syntax of PLA} of the syntax of $PLA$) to get a similar formula
for any $C \in (0, 1)$.

We simply let $\varphi_0(x, y)$ be the formula $x = y$.
Suppose that $\varphi_k(x, y) \in PLA(\sigma)$ is such that, for all $a, b \in A$, if $a = b$ or if
$I(a) \neq \es$ and $I(b) \neq \es$, then
$\mcA(\varphi_k(a, b)) = s_k(a, b)$.
Then let $\psi(x, y, u, v)$ be the formula $E(u, x) \wedge E(v, y)$ and let
$\varphi'_k(x, y, u, v)$ be $\varphi_k(u, v) \wedge \psi(x, y, u, v)$.
Define $\varphi_{k+1}(x, y)$ to be the formula
\[
x = y \ \vee \ 
\mr{cam}^*(\varphi'_k(x, y, u, v), \varphi'_k(x, y, u, u), \psi(x, y, u, v), \psi(x, y, u, u) : u, v : 
p^=(x, y, u, v))
\]
where $\mr{cam}^*$ is the admissible aggregation function defined in 
Example~\ref{conditional arithmetic mean} and 
$p^=(x, y, u, v)$ expresses that all of $x, y, u, v$ are different.
Then, if $a = b$ or if 
$I(a) \neq \es$ and $I(b) \neq \es$ we have
$\mcA(\varphi_{k+1}(a, b)) = s_{k+1}(a, b)$.
}\end{exam}

\section{Admissibility and the main result}\label{Admissibility and the main results}

\noindent
We begin by considering the condition on aggregation functions, admissibility, that will allow us to asymptotically
eliminate them in the context of distributions induced by lifted Bayesian networks.

\subsection{Admissibility}

Our main result tells that `admissible' aggregation functions can be asymptotically eliminated from $PLA$-formulas.
Admissibility is a kind of continuity condition and to define it we will use the notion of 
convergence testing sequence. Informally speaking, an infinite sequence $\bar{r}_n \in [0, 1]^{<\omega}$, $n \in \mbbN$,
is convergence testing if $|\bar{r}_n| < |\bar{r}_{n+1}|$ for all $n$ and if there are $k \in \mbbN^+$
and $c_1, \ldots, c_k, \alpha_1, \ldots, \alpha_k \in [0, 1]$ such that, as $n\to\infty$, 
every entry of $\bar{r}_n$ is ever closer to one of $c_1, \ldots, c_k$ and, for $i = 1, \ldots, k$,
the proportion of entries in $\bar{r}_n$
that are close to $c_i$ is ever closer to $\alpha_i$.
Our definition of convergence testing sequence is similar in spirit to a stronger notion with the same name used by
Jaeger \cite{Jae98a}. The corresponding notion in \cite{Jae98a} is stronger than ours because it adds a requirement that 
``accumulation'' around certain points happens with exponential speed.

\begin{defin}\label{definition of convergence testing}{\rm 
A sequence $\bar{r}_n \in [0, 1]^{<\omega}$, $n \in \mbbN$,  is called {\em convergence testing} for parameters 
$c_1, \ldots, c_k \in [0,1]$ and $\alpha_1, \ldots \alpha_k  \in  [0,1]$ if the following hold, 
where $r_{n,i}$ denotes the $i$th entry of $\bar{r}_n$:
\begin{enumerate}
\item $|\bar{r}_n| < |\bar{r}_{n+1}|$ for all $n \in \mbbN$.
\item For every disjoint family of open intervals $I_1, \ldots I_k \subseteq [0,1]$ such that $c_i \in I_i$ for each $i$, 
there is an $ N \in \mbbN$ such that $\mathrm{rng}(\bar{r}_n) \subseteq \bigcup\limits_{j=1}^{k} I_j$ for all $n \geq N$, 
and for every $j \in \{1, \ldots, k \}$, 
\[
\lim\limits_{n \rightarrow \infty} \frac{\left| \{ i \leq |\bar{r}_n| : r_{n,i} \in I_j \} \right| }{|\bar{r}_n|} = \alpha_j
\]
\end{enumerate}   

More generally, a sequence of $k$-tuples of sequences 
$(\bar{r}_{1, n}, \ldots, \bar{r}_{k, n}) \in  \big([0, 1]^{<\omega}\big)^k$, $n \in \mbbN$,  
is called {\em convergence testing} for parameters $c_{i,j} \in [0,1]$ and $\alpha_{i,j} \in [0,1]$, 
where $i \in \{1, \ldots, k\}$, $j \in \{ 1, \ldots, m_i \}$ and $m_1, \ldots m_k \in \mbbN^+$, if  for every fixed $i \in \{1, \ldots, k \}$
the sequence $\bar{r}_{i, n}$, $n \in \mbbN$, is convergence testing for $c_{i,1}, \ldots, c_{i, m_i}$, and $\alpha_{i,1}, \ldots, \alpha_{i, m_i}$.
}\end{defin}

\noindent
Roughly speaking, a function is {\em admissible} if it is continuous for all sequences of fixed length 
$n \in \mathbb{N}$ and continuous on convergence testing sequences for $n \rightarrow \infty$.
More precisely we have:

\begin{defin} \label{alternative definition of admissible function}{\bf (Admissibility)} {\rm 
An aggregation function $F : \big([0, 1]^{<\omega}\big)^k \to [0, 1]$ is called {\em admissible} if the following two conditions hold:
\begin{enumerate}
\item For all $n_1, \ldots, n_k \in \mbbN^+$, $F$ is continuous on the set $[0, 1]^{n_1} \times \dots \times [0, 1]^{n_k}$.
\item For all convergence testing sequences of tuples
$(\bar{r}_{1, n}, \ldots, \bar{r}_{k, n}) \in  \big([0, 1]^{<\omega}\big)^k$, $n \in \mbbN$,
and $(\bar{\rho}_{1, n}, \ldots, \bar{\rho}_{k, n}) \in  \big([0, 1]^{<\omega}\big)^k$, $n \in \mbbN$,
with the same parameters $c_{i,j} \in [0, 1]$ and $\alpha_{i,j} \in (0, 1]$, 
$\underset{n \rightarrow \infty}{\lim}  |F(\bar{r}_{1, n}, \ldots, \bar{r}_{k, n}) - F(\bar{\rho}_{1, n}, \ldots, \bar{\rho}_{k, n})| = 0$.
\end{enumerate}
}\end{defin}

\noindent
Observe that in the definition of admissibility we require that all $\alpha_{i, j}$ are nonzero.
It is straightforward to verify that Noisy-or is not admissible, but we have:

\begin{prop}\label{some common functions are admissible}
The functions {\rm am} (arithmetic mean), {\rm gm} (geometric mean), {\rm max} and {\rm min} are admissible.
\end{prop}

\noindent
{\bf Proof.}
All functions are clearly continuous on $[0,1]^n$, for every $n$, as can be seen directly from their definition.
So we proceed to show that they are compatible with convergence testing sequences as demanded by Condition (2) of admissibility. 
So let $\bar{r}_n$, $n \in \mbbN$, be convergence testing with parameters 
$c_{1}, \ldots, c_k \in [0, 1]$ and $\alpha_1, \ldots,\alpha_k \in (0, 1]$. 
Let $F$ be the arithmetic mean.
Then $\underset{n \rightarrow \infty}{\lim} F(\bar{r}_n) = \alpha_1 c_1 + \dots + \alpha_k c_k.$ 
Indeed, for any sufficiently small $\delta > 0$ there is an $N \in \mathbb{N}$ such that for all $n > N$,  

\begin{align}\label{rngcond}
\mathrm{rng} (\bar{r}_n) \subseteq \bigcup_{i=1}^k(c_i - \delta, c_i + \delta)
\end{align}
 and 
 \begin{align}\label{propcond}
 \frac{\left| \{ i \leq |\bar{r}_n| : r_{n} \in I_i \} \right| }{|\bar{r}_n|} \in (\alpha_i- \delta, \alpha_i + \delta).
 \end{align}
 But then 
\[
(\alpha_1- \delta) (c_1 - \delta) + \dots + (\alpha_k - \delta) (c_k - \delta) < 
F(\bar{r}_n) < (\alpha_1 + \delta) (c_1 + \delta) + \dots + (\alpha_k + \delta) (c_k + \delta),
\]
with both bounds converging to $\alpha_1 c_1 + \dots + \alpha_k c_k$ as $\delta \rightarrow 0$. 
If $F$ is ${\rm max}$ or ${\rm min}$, we can use an analogous argument, with 
\[
{\rm max}(c_1, \dots, c_k) - \delta < {\rm max}(\bar{r}_n) < {\rm max}(c_1, \dots, c_k) + \delta
\]
and 
\[
{\rm min}(c_1, \dots, c_k) - \delta < {\rm min}(\bar{r}_n) < {\rm min}(c_1, \dots, c_k) + \delta.
\]
For $F$ the geometric mean we obtain $\underset{n \rightarrow \infty}{\lim} F(\bar{r}_n) = \prod_{i =1}^{k} c_i^{\alpha_i}$. 
Indeed, for any $\delta > 0$ choose $N$ such that~(\ref{rngcond}) and~(\ref{propcond}) hold for all $n > N$. 
Then for all $n > N$, 
$\prod_{i =1}^{k} (c_i - \delta)^{\alpha_i + \delta} < F(\bar{r}_n) < 
\prod_{i =1}^{k} (c_i + \delta)^{\alpha_i - \delta}$, 
with both bounds converging to $\prod_{i =1}^{k} c_i^{\alpha_i}$. 
\hfill $\square$
\\

\begin{exam}\label{non-unary admissible aggregation functions} {\bf (Non-unary aggregation functions)}
{\rm Here are some examples, besides `cam' and `$\mr{cam}^*$' from
Examples~\ref{simple case of conditional arithmetic mean} 
and~\ref{conditional arithmetic mean},
of aggregation functions that take two or more sequences as input.

Define $F: \big([0, 1]^{<\omega}\big)^2 \to [0, 1]$ by
$F(\bar{r}, \bar{\rho}) = |\mr{am}(\bar{r}) - \mr{am}(\rho)|$. It can be used
when some quantity is influenced by the imbalance of (the means of) two other quantities.
Since `$\mr{am}$' is admissible and $|x - y|$ is uniformly continuous on $\mbbR^2$ it follows that 
$F$ is admissible.

For another example, let $S(x) : \mbbR \to [0, 1]$ be the {\em sigmoid} function, that is,
$S(x) = (1 + e^{-x})^{-1}$.
Let $k > 1$ and let the ``weights'' $w_1, \ldots, w_k \in [0, 1]$ be such that $w_1 + \ldots + w_k = 1$.
Define $G: \big([0, 1]^{<\omega}\big)^k \to [0, 1]$ by 
$G(\bar{r}_1, \ldots, \bar{r}_k) = S\big(\sum_{i=1}^k w_i \cdot \mr{am}(\bar{r}_i)\big)$.
$G$ is used in the context of {\em Domain-size-Aware Relational Logistic Regression} models \cite{Wei21a} and
it is admissible because `$\mr{am}$' is admissible and $S$ is uniformly continuous.

As a third example, which is not an ``arithmetic combination'' of unary aggregation functions (such as am or gm) we have
the pseudometric $\mu_1^u$ on $[0, 1]^{<\omega}$, described in Definition~\ref{definition of the metric} below,
which is a binary aggregation function.
}\end{exam}

\noindent
In Examples~\ref{simple case of conditional arithmetic mean}
and~\ref{conditional arithmetic mean}
we considered the ``conditional arithmetic means'' cam and $\mr{cam}^*$.
The next proposition tells that they are indeed admissible.

\begin{prop}\label{cam is admissible}
The aggregation functions cam and $\mr{cam}^*$ are admissible.
\end{prop}

\noindent
{\bf Proof.}
We first consider cam.
Recall its  definition: 
$\mr{cam}(\bar{p}, \bar{r}) = 0$ if  $\bar{p}$ contains only zeros, and otherwise 
\begin{align*}
\mr{cam}(\bar{p}, \bar{r})  =  \frac{\mr{am}(\bar{p})}{
\max(\mr{am}(\sqrt{\bar{p}}), \mr{am}(\bar{r}))}.
\end{align*}
We begin by verifying that condition~(1) in the definition of admissible aggregation function
is satisfied. In other words, we need to check that, for all $n, m \in \mbbN^+$, cam is continuous at every point 
$\bar{p}\bar{r} \in [0, 1]^n \times [0, 1]^m$.
By the definition of cam this is clear for all points $\bar{p}\bar{r}$ such that $\bar{p}$ does not contain only
zeros. So suppose that $\bar{p}_0 \in [0, 1]^n$ contains only zeros and that $\bar{r}_0$ is any point in 
$[0, 1]^m$.
It is straightforward to verify that
\begin{equation}\label{upper bound of cam}
\text{if $\bar{p}$ does not consist of only zeros, then } 
\mr{cam}(\bar{p}, \bar{r}) \leq \frac{\mr{am}(\bar{p})}{\mr{am}(\sqrt{\bar{p}})}.
\end{equation}
So if $\bar{p}_n$, respectively $\bar{r}_n$, $n \in \mbbN^+$, are sequences that tend to $\bar{p}_0$,
respectively $\bar{r}_0$, then the
right hand side of~(\ref{upper bound of cam}) 
tends to 0, and therefore $\mr{cam}(\bar{p}_n, \bar{r}_n)$ tends to 0.

Now we verify condition~(2) in the definition of admissibility.
Suppose that $\bar{p}_n$, $n \in \mbbN^+$, is a convergence testing sequence
with parameters $a_1, \ldots, a_k \in [0, 1]$ and $\alpha_1, \ldots, \alpha_k \in (0, 1]$ and that
$\bar{r}_n$, $n \in \mbbN^+$, is a convergence testing sequence with parameters 
$b_1, \ldots, b_l \in [0, 1]$ and $\beta_1, \ldots, \beta_l \in (0, 1]$.
It suffices to show that $\lim_{n\to\infty}\mr{cam}(\bar{p}_n, \bar{r}_n)$ exists and depends
only on the parameters.

First suppose that all $a_i$ are zero. 
Then $\lim_{n\to\infty}\max(\bar{p}_n) = 0$ and it follows from~(\ref{upper bound of cam})
that $\lim_{n\to\infty}\mr{cam}(\bar{p}_n, \bar{r}_n) = 0$.

Next, suppose that at least one $a_i$ is positive.
In this case 
\[
\lim_{n\to\infty}\mr{cam}(\bar{p}_n, \bar{r}_n) = 
\frac{\sum_{i=1}^k \alpha_i a_i}
{\max\big[ \sum_{i=1}^k \alpha_i \sqrt{a_i}, \ \sum_{i=1}^l \beta_i b_i \big]}
\]
which is well defined because $\sum_{i=1}^k \alpha_i \sqrt{a_i} > 0$ (as all $\alpha_i > 0$).

Now we consider the aggregation function $\mr{cam}^*$.
Let $f^{-1}$ denote the inverse of the function $f(x) = x(x-1)$ on the positive reals.
Let $\lambda(\bar{p}, \bar{r}) = 1 - \max(\bar{p}\bar{r})(1 - \frac{1}{m})$
where $m = f^{-1}(|\bar{p}|)$. 
Recall that, for $\bar{p}, \bar{q}, \bar{r}, \bar{s} \in [0, 1]^{<\omega}$,
\[
\text{$\mr{cam}^*(\bar{p}, \bar{q}, \bar{r}, \bar{s}) = 0$ if $\bar{p}$ and $\bar{q}$ contain only zeros,}
\]
and otherwise
\begin{align}
&\mr{cam}^*(\bar{p}, \bar{q}, \bar{r}, \bar{s}) = \nonumber \\
&\frac{\max(\bar{p} \bar{r}) \mr{am}(\bar{p}) + 
\lambda(\bar{p}, \bar{r})\mr{am}(\bar{q})}
{\max\big[ \max(\bar{p} \bar{r}) \mr{am}(\sqrt{\bar{p}}) + 
\lambda(\bar{p}, \bar{r})\mr{am}(\sqrt{\bar{q}}), \ 
\max(\bar{p}\bar{r})\mr{am}(\bar{r}) + \lambda(\bar{p}, \bar{r})\mr{am}(\bar{s})\big] }. 
\label{definition of cam*}
\end{align}

Before verifying conditions~(1) and~(2) in the definition of admissibility, we observe that
if it is not the case that both $\bar{p}$ and $\bar{q}$ are constantly zero, then the following holds,
which straightforward to verify:
\begin{equation}\label{upper bound on cam*}
\mr{cam}^*(\bar{p}, \bar{q}, \bar{r}, \bar{s}) \ \leq \ 
\frac{\mr{am}(\bar{p})}{\mr{am}(\sqrt{\bar{p}})} + 
\frac{\mr{am}(\bar{q})}{\mr{am}(\sqrt{\bar{q}})}.
\end{equation}

We first show that, for all $n_1, n_2, n_3, n_4 \in \mbbN^+$,
$\mr{cam}^*$ is continuous at every point 
$(\bar{p}_0, \bar{q}_0, \bar{r}_0, \bar{s}_0) \in 
[0, 1]^{n_1}  \times [0, 1]^{n_2} \times [0, 1]^{n_3} \times [0, 1]^{n_4}$.
Note that the number $m$ in the expression $\lambda(\bar{p}_0, \bar{r}_0)$ 
depends only on $|\bar{p}_0| = n_1$.
As $\mr{cam}^*$ is constructed by composing arithmetic operations, max, the function $\mr{am}$ and using the
square root, its continuity at $(\bar{p}_0, \bar{q}_0, \bar{r}_0, \bar{s}_0)$
is clear whenever $\bar{p}_0$ and $\bar{q}_0$ do not consist only of zeros.
So now suppose that both $\bar{p}_0$ and $\bar{q}_0$ consist only of zeros,
so $\mr{cam}^*(\bar{p}_0, \bar{q}_0, \bar{r}_0, \bar{s}_0) = 0$.
If $(\bar{p}, \bar{q}, \bar{r}, \bar{s})$ approaches $(\bar{p}_0, \bar{q}_0, \bar{r}_0, \bar{s}_0)$,
then $\bar{p}$ and $\bar{q}$ approach sequences which are constantly zero, so
the right hand side of~(\ref{upper bound on cam*}) approaches 0 and hence
$\mr{cam}^*(\bar{p}, \bar{q}, \bar{r}, \bar{s})$ approaches 0.

Next, we must show that~(2) in the definition of admissibility holds.
Suppose that $(\bar{p}_n, \bar{q}_n, \bar{r}_n, \bar{s}_n) \in \big([0, 1]^{<\omega}\big)^4$,
$n \in \mbbN^+$, is a convergence testing sequence with parameters 
\begin{align*}
&a_1, \ldots, a_{k_1}, b_1, \ldots, b_{k_2}, c_1, \ldots, c_{k_3}, d_1, \ldots, d_{k_4} \in [0, 1] \text{ and} \\
&\alpha_1, \ldots, \alpha_{k_1}, \beta_1, \ldots, \beta_{k_2}, \gamma_1, \ldots, \gamma_{k_3},
\delta_1, \ldots, \delta_{k_4} \in (0, 1].
\end{align*}
It suffices to show that 
\begin{equation}\label{limit of cam*}
\lim_{n\to\infty}\mr{cam}^*(\bar{p}_n, \bar{q}_n, \bar{r}_n, \bar{s}_n)
\end{equation}
exists and depends only on the parameters.
Note that  $\lim_{n\to\infty}\mr{am}(\bar{p}_n) = \sum_{i=1}^{k_1} \alpha_i a_i$,
$\lim_{n\to\infty}\mr{am}(\sqrt{\bar{p}_n}) = \sum_{i=1}^{k_1} \alpha_i \sqrt{a_i}$, and similarly
for the other sequences.
Note also that 
\[
\lim_{n\to\infty}\max(\bar{p}_n\bar{r}_n) = 
\max\{a_1, \ldots, a_{k_1}, c_1, \ldots, c_{k_3}\}.
\]
Moreover, as $n \to \infty$ we have $m \to \infty$.

It follows that both the enumerator and denominator in the expression~(\ref{definition of cam*})
of \\
$\mr{cam}^*(\bar{p}_n, \bar{q}_n, \bar{r}_n, \bar{s}_n)$ converges as $n \to \infty$.
Hence, if the denominator converges to a positive number then 
$\lim_{n\to\infty}\mr{cam}^*(\bar{p}_n, \bar{q}_n, \bar{r}_n, \bar{s}_n)$ exists.
We make case distinctions with respect to the parameters.

Suppose that at least one $a_i$ is positive.
Then $\max(\bar{p}_n \bar{r}_n)$ converges to a positive number and
$\mr{am}(\sqrt{\bar{p}_n})$ converges to a positive number (namely $\sum_i \alpha_i \sqrt{a_i}$),
so the denominator in~(\ref{definition of cam*}) converges to a positive number.

If some $c_i$ is positive then 
$\max(\bar{p}_n \bar{r}_n)\mr{am}(\bar{r}_n)$ converges to a positive number and hence
the denominator in~(\ref{definition of cam*}) converges to a positive number.

Suppose that all $a_i$ and all $c_i$ are zero and some $d_i$ is positive.
Then $\max(\bar{p}_n \bar{r}_n)$ converges to 0 and hence
$\lambda(\bar{p}_n, \bar{r}_n)$ converges to a positive number.
Also, $\mr{am}(\bar{s}_n)$ converges to $\sum_i \delta_i d_i$ which is positive
since all $\delta_i$ are positive. 
Hence $\lambda(\bar{p}_n, \bar{r}_n)\mr{am}(\bar{s}_n)$ converges to a positive number
and thus the same holds for the denominator in~(\ref{definition of cam*}).

Suppose that all $a_i$, all $c_i$ and all $d_i$ are zero, but at least one $b_i$ is positive.
Then 
\[
\lim_{n\to\infty}\mr{cam}^*(\bar{p}_n, \bar{q}_n, \bar{r}_n, \bar{s}_n) = 
\lim_{n\to\infty}\frac{\mr{am}(\bar{q}_n)}{\mr{am}(\sqrt{\bar{q}_n})} = 
\frac{\sum_i \beta_i b_i}{\sum_i \beta_i \sqrt{b_i}}
\]
where the denominator is positive since all $\beta_i$ are positive.

Finally, suppose that all $a_i$, all $b_i$, all $c_i$ and all $d_i$ are zero.
Then it follows from~(\ref{upper bound on cam*}) that 
$\lim_{n\to\infty}\mr{cam}^*(\bar{p}_n, \bar{q}_n, \bar{r}_n, \bar{s}_n) = 0$.
\hfill $\square$
\\

\subsection{Noncriticallity}

Let $\sigma$ be a finite relational signature.
The main result uses the assumption that every aggregation formula of the lifted Bayesian network $\mbbG$ (for $\sigma$) used to define
probability distributions is noncritical with respect to the network.
The notion of noncritical $CPL(\sigma)$-formula defined in \cite{Kop20} uses the notion of $m$-critical number, where $m \in \mbbN$.
The notion of $m$-critical number in \cite[Definition~4.29]{Kop20} is quite technical and is embedded in the proof of the main results
of \cite{Kop20}.
But it follows from Lemma~4.12 and
Definitions~4.18, 4.22,
and~4.29 in \cite{Kop20},
that if $\alpha \in \mbbR$ is $m$-critical (with respect to a lifted Bayesian network $\mbbG$ for $\sigma$)
in the sense of  \cite[Definition~4.29]{Kop20}, then it can be generated,
using the operations addition, multiplication and division from 
the set of numbers
\[
S(\mbbG) = \{0, 1\} \cup \{\mu(R \ | \ \chi_{R, i}) : R \in \sigma\} \cup \{1 - \mu(R \ | \ \chi_{R, i}) : R \in \sigma\}
\]
where $\mu(R \ | \ \chi_{R, i})$ is the number associated to $\mbbG$ in
Definition~\ref{definition of BN}.
In fact, if $\alpha$ is $m$-critical in the sense of \cite[Definition~4.29]{Kop20},
then $\alpha$ can be generated from $S(\mbbG)$ with at most $16^{l^l \cdot |\sigma|}$ applications of the operations
addition, multiplication and division, where $l$ is the sum of $m$ and the maximal arity of the relation symbols in $\sigma$.
The number $16^{l^l \cdot |\sigma|}$ is a very crude upper bound based on considering 
the possible atomic $\sigma$-types
in $m$ variables and the definitions and result from \cite{Kop20} mentioned above.
To describe noncritical formulas more easily we will use the following definition.

\begin{defin}\label{definition of quantifier parameter} {\rm
Let $\varphi \in CPL(\sigma)$.
A real number $r$ is a {\em quantifier parameter of $\varphi$} if 
$\varphi$ has a subformula of the form
\begin{align*}
\Big( r + \| \varphi \ |  \ \psi \|_{\bar{y}} \ \geq \ 
\| \theta \ |  \ \tau \|_{\bar{y}} \Big) 
\ \ \text{ or} \ \ 
\Big( \| \varphi \ |  \ \psi \|_{\bar{y}} \ \geq \ 
\| \theta \ |  \ \tau \|_{\bar{y}} + r \Big).
\end{align*}
}\end{defin}
If a formula $\varphi \in CPL(\sigma)$ is critical with respect to a lifted Bayesian network $\mbbG$,
in the sense of \cite[Definition~4.30]{Kop20},
then $\varphi$ has a quantifier parameter $r$ such that $r = \alpha - \beta$ and both $\alpha$ and $\beta$
are $m$-critical in the sense of \cite[Definition~4.29]{Kop20} where $m$ is the sum of the number of free variables in $\varphi$ 
and the quantifier-rank of $\varphi$ (in the sense of \cite[Definition~3.7]{Kop20}).
It follows from the discussion above that if $\varphi$ is critical with respect to $\mbbG$ in the sense of \cite[Definition~4.30]{Kop20},
then $\varphi$ has a quantifier parameter $r$ such that $r = \alpha - \beta$ and both $\alpha$ and $\beta$ can be
generated from $S(\mbbG)$, as defined above, by at most $16^{l^l \cdot |\sigma|}$ applications of the operations addition, multiplication and division, where $l$ is the sum of the length of $\varphi$ (as a string of symbols) and the maximal arity of the relation symbols
in $\sigma$.

In order to avoid the technicalities involved in \cite[Definitions~4.29 and~4.30]{Kop20}
we will define a notion of noncritical formula which is somewhat stronger than
the corresponding notion in \cite[Definition~4.30]{Kop20} but still interesting, we think, 
since for any $k$, the set of numbers which can be generated from $S(\mbbG)$ with at most $k$ applications of 
addition, multiplication and division is finite, so this set can be avoided by ``moving'' a tiny bit up or down in $\mbbR$.

\begin{defin}\label{criticality of CPL-formulas} {\bf (Noncritical formula with respect to $\mbbG$)} {\rm
Let $\mbbG$ be a lifted Bayesian network for $\sigma$. 
We call a formula $\varphi \in CPL(\sigma)$ {\em potentially critical with respect to $\mbbG$} if it has a quantifier parameter
$r$ such that $r = \alpha - \beta$ and both $\alpha$ and $\beta$
can be generated from $S(\mbbG)$ (as defined above) 
by at most $16^{l^l \cdot |\sigma|}$ applications of the operations 
addition, multiplication and division, where $l$ is the sum of the length of $\varphi$ (as a string of symbols) and the maximal arity of the relation symbols in $\sigma$.
Otherwise we call $\varphi$ {\em noncritical with respect to $\mbbG$}. 
}\end{defin}

\noindent
It follows that every first-order formula is noncritical with respect to any lifted Bayesian network.

\subsection{The main result}

\begin{theor}\label{elimination of aggregation functions for an LBN} 
{\bf (Asymptotic elimination of admissible aggregation functions)} 
Let $\sigma$ be a finite relational signature and let 
$\mbbG$ be a lifted Bayesian network for $\sigma$
such that every aggregation formula of $\mbbG$ is noncritical with respect to $\mbbG$.
If $\varphi(\bar{x}) \in PLA(\sigma)$ and all aggregation functions in $\varphi$ are admissible, 
then $\varphi(\bar{x})$ is asymptotically equivalent to a basic probability formula with respect to 
the sequence of probability distributions induced by $\mbbG$.
\end{theor}

\noindent
The proof of 
Theorem~\ref{elimination of aggregation functions for an LBN}
is carried out in 
Section~\ref{asymptotic elimination of aggregation functions}.
First we derive a corollary.

\begin{cor}\label{corollary to main results} {\bf (Convergence of probability)}
Let $\mbbG$ be a lifted Bayesian network over a finite relational signature $\sigma$ 
such that every aggregation formula of $\mbbG$ is noncritical with respect to $\mbbG$.
Let $(\mbbP_n : n \in \mbbN^+)$ be the sequence of probability distributions induced by $\mbbG$.
If $\varphi(\bar{x}) \in PLA(\sigma)$ has only admissible aggregation functions 
then there are $c_1, \ldots, c_k \in [0, 1]$, depending only on $\varphi$ and $\mbbG$, such that
for every $m \in \mbbN^+$, every $\bar{a} \in [m]^{|\bar{x}|}$ and every $\varepsilon > 0$,
\begin{align*}
&\lim_{n\to\infty} \mbbP_n\big(\big\{ \mcA \in \mbW_n : 
\mcA(\varphi(\bar{a})) \in \bigcup_{i=1}^k[c_i - \varepsilon, c_i + \varepsilon] \big\} \big) \ = \ 1\\ 
&\text{and for all $i = 1, \ldots, k$ } \\
&\mbbP_n\big(\big\{\mcA \in \mbW_n : |\mcA(\varphi(\bar{a})) - c_i| < \varepsilon \big\}\big) 
\text{ converges as } n\to\infty.
\end{align*}
\end{cor}

\noindent
{\bf Proof.}
Let $\mbbG$, $\mbbP = (\mbbP_n : n \in \mbbN^+)$ and $\varphi(\bar{x})$ be as assumed.
By 
Theorem~\ref{elimination of aggregation functions for an LBN},
there is a basic probability formula $\psi(\bar{x})$ which is asymptotically equivalent to $\varphi(\bar{x})$
with respect to $\mbbP$.
Then $\psi(\bar{x})$ has the form $\bigwedge_{i=1}^k (\psi_i(\bar{x}) \to c_i)$ where, for each $i$,
$c_i \in [0, 1]$ and $\psi_i(\bar{x})$ is a
conjunction of first-order literals. 
Without loss of generality we can assume that each $\psi_i$ is the conjunction of all 
formulas in a complete atomic $\sigma$-type.
Note that for every $\mcA \in \mbW_n$ and every $\bar{a} \in [n]^{|\bar{x}|}$ we have 
$\mcA(\psi(\bar{a})) \in \{c_1, \ldots, c_k\}$ and $\mcA(\psi(\bar{a})) = c_i$ if $\mcA \models \psi_i(\bar{a})$.
Let $c \in \{c_1, \ldots, c_k\}$ and suppose that $i_1, \ldots, i_t$ enumerates all $i$ such that $c_i = c$.
Then 
\[
\mbbP_n\big(\{\mcA \in \mbW_n : \mcA(\psi(\bar{a})) = c\}\big) = \mbbP_n\big(\bigvee_{j=1}^t \psi_{i_j}(\bar{a})\big).
\]
By Proposition~\ref{convergence and saturation if associated formulas are bpf}
below, 
it follows that the above probability converges as $n\to\infty$. (Moreover, the number to which it converges
depends only on $\psi$ and $\mbbG$, according to the same proposition.)
Since $\varphi(\bar{x})$ and $\psi(\bar{x})$ are asymptotically equivalent with respect to $\mbbP$
the conclusions of the corollary follow.
\hfill $\square$
\\

\noindent

\section{Asymptotic elimination of aggregation functions}\label{asymptotic elimination of aggregation functions}

\noindent
In this section we prove 
Theorem~\ref{elimination of aggregation functions for an LBN}.
Its proof is concluded by 
Corollary~\ref{elimination of aggregation functions if saturation conditions hold}.
The definition of admissible aggregation function given above 
(Definition~\ref{alternative definition of admissible function})
is relatively intuitive and was convenient for proving 
Proposition~\ref{some common functions are admissible}.
But in the proofs of this section another characterization of admissibility is needed.
The next subsection shows that admissibility is equivalent to a condition which we call ``admissibility sensu novo''.

\subsection{An alternative characterization of admissibility}\label{An alternative characterization of admissibility}

In order to formulate the other characterization of admissibility we  need to relate each $\bar{r} \in [0, 1]^{<\omega}$
to a specific function from $[0, 1]$ to $[0, 1]$, and we need to define a couple of pseudometrics on $[0, 1]^{<\omega}$.

\begin{defin}\label{definition of associated function}{\bf (Functional representations of sequences)} {\rm
Let $n \in \mbbN^+$ and let $\bar{r} = (r_1, \ldots, r_n) \in [0, 1]^n$. We will associate a function from $[0, 1]$ to $[0, 1]$ with $\bar{r}$ in
two different ways, one way where the order of the entries in $\bar{r}$ matters and one in which the order does not influence the associated function.
\begin{enumerate}
\item Define $\mff_{\bar{r}}$, which we call the {\em ordered functional representation of $\bar{r}$}, as follows: 
For every $a \in [0, 1/n)$, let $\mff_{\bar{r}}(a) = r_1$, 
for every $i = 1, \ldots, n-1$ and every $a \in [i/n, (i+1)/n)$, let $f(a) = r_{i+1}$ and finally let $f(1) = r_n$.

\item Define $\mfg_{\bar{r}}$, which we call the {\em unordered functional representation of $\bar{r}$}, as follows:
Let $\bar{\rho} = (\rho_1, \ldots, \rho_n)$ be a reordering of $\bar{r}$ such that, for all $i = 1, \ldots, n-1$, 
$\rho_i \leq \rho_{i+1}$ and let $\mfg_{\bar{r}} = \mff_{\bar{\rho}}$. 
\end{enumerate}
}\end{defin}

\noindent
In both the ordered and unordered case we consider two different ways to measure how 
different two sequences $\bar{r}, \bar{\rho} \in [0, 1]^{<\omega}$ are (where the sequences may have different lengths).

\begin{defin}\label{definition of the metric}{\bf (Pseudometrics on sequences)} {\rm
\begin{enumerate}

\item First we recall the $L_1$ and $L_\infty$ norms: for every (bounded and integrable) $f : [0, 1] \to \mbbR$ they are defined as
\[
\| f \|_1 = \int_{[0,1]} | f(x) |dx \qquad \text{ and} \qquad
\| f \|_\infty = \sup\{|f(a)| : a \in [0, 1]\}.
\]
\item For $\bar{r}, \bar{\rho} \in [0, 1]^{<\omega}$ we define 
\begin{align*}
&\mu_1^o(\bar{r}, \bar{\rho}) =  \|\mff_{\bar{r}} - \mff_{\bar{\rho}}\|_1, \\
&\mu_1^u(\bar{r}, \bar{\rho}) =  \|\mfg_{\bar{r}} - \mfg_{\bar{\rho}}\|_1, \\
&\mu_\infty^o(\bar{r}, \bar{\rho}) =  \|\mff_{\bar{r}} - \mff_{\bar{\rho}}\|_\infty, \\
&\mu_\infty^u(\bar{r}, \bar{\rho}) =  \|\mfg_{\bar{r}} - \mfg_{\bar{\rho}}\|_\infty.
\end{align*}

\item Let $\mu$ denote any one of the four functions in the previous part.
For arbitrary $k > 1$ we can define a function on $\big([0, 1]^{<\omega}\big)^k$, also denoted $\mu$ (to avoid making notation more complicated),
as follows: For all $(\bar{r}_1, \ldots, \bar{r}_k), (\bar{r}'_1, \ldots, \bar{r'}_k) \in \big([0, 1]^{<\omega}\big)^k$
let 
\[
\mu\big((\bar{r}_1, \ldots, \bar{r}_k), (\bar{r}'_1, \ldots, \bar{r'}_k)\big) = 
\max\big(\mu(\bar{r}_1, \bar{r}'_1), \ldots, \mu(\bar{r}_k, \bar{r}'_k)\big)
\]
\end{enumerate}
}\end{defin}

\noindent
From well-known results in analysis it follows that the functions 
$\mu_1^o, \mu_1^u, \mu_\infty^o$ and $\mu_\infty^u$ are symmetric and satisfy the triangle inequality so they are
pseudometrics on $[0, 1]^{<\omega}$, and the same is true for the variants defined on $\big([0, 1]^{<\omega}\big)^k$ with $k > 1$.
We may have $\mu_1^u(\bar{r}, \bar{\rho}) = 0$ and $\bar{r} \neq \bar{\rho}$,
for example if $\bar{r} = (0, 1/2, 1)$ and $\bar{\rho} = (0, 0, 1/2, 1/2, 1, 1)$, so $\mu_1^u$ is not a metric. 
Similar examples show that the $\mu_1^o, \mu_\infty^o$ and $\mu_\infty^u$ are not metrics.
Also note that for all $\bar{r}, \bar{\rho} \in [0, 1]^{<\omega}$, $\mu_1^o(\bar{r}, \bar{\rho}) \leq 1$ and similarly for
$\mu_1^u, \mu_\infty^o$ and $\mu_\infty^u$.
If $\bar{r} = (r_1, \ldots, r_n)$ and $\bar{\rho} = (\rho_1, \ldots, \rho_n)$ have the same length $n$, then we simply have
$\mu_\infty^o(\bar{r}, \bar{\rho}) = \max\{|r_i - \rho_i| : i = 1, \ldots, n\}$, but in general we can not reduce 
$\mu_\infty^o(\bar{r}, \bar{\rho}), \mu_\infty^u(\bar{r}, \bar{\rho}), \mu_1^o(\bar{r}, \bar{\rho})$ or
$\mu_1^u(\bar{r}, \bar{\rho})$ to computing the maximal coordinatewise difference.

\begin{defin}\label{definition of continuous aggregation function}
{\rm
Let $F : \big([0, 1]^{<\omega}\big)^k \to [0, 1]$ be an aggregation function and let 
$\mu$ be any of the the pseudometrics defined in Definition~\ref{definition of the metric}.
Also let $X \subseteq \big([0, 1]^{<\omega}\big)^k$.
We say that $F$ is {\em asymptotically uniformly continuous (with respect to $\mu$) on $X$}
if for every $\varepsilon > 0$ there are $n$ and $\delta > 0$ such that 
if $(\bar{r}_1, \ldots, \bar{r}_k), (\bar{\rho}_1, \ldots, \bar{\rho}_k) \in X$,
$|\bar{r}_i|, |\bar{\rho}_i| \geq n$ for all $i$ and 
$\mu\big((\bar{r}_1, \ldots, \bar{r}_k), (\bar{\rho}_1, \ldots, \bar{\rho}_k)\big) < \delta$, then 
$\big|F(\bar{r}_1, \ldots, \bar{r}_k) - F(\bar{\rho}_1, \ldots, \bar{\rho}_k)\big| < \varepsilon$.
}\end{defin}

\begin{defin}\label{definition of admissible function}{\bf (Alternative characterization of admissibility)} 
{\rm
An aggregation function $F : \big([0, 1]^{<\omega}\big)^k \to [0, 1]$ is called {\em admissible sensu novo} if the following two conditions hold:
\begin{enumerate}
\item For all $m_1, \ldots, m_k \in \mbbN^+$, all $c_{i, j} \in [0, 1]$ and $\alpha_{i, j} \in (0, 1]$, 
for $i = 1, \ldots, k$ and $j = 1, \ldots, m_i$,
and all sufficiently small $\delta > 0$,
$F$ is asymptotically uniformly continuous with respect to $\mu_1^u$ on $X_1 \times \ldots \times X_k$ where, for each $i = 1, \ldots, k$,
\begin{align*}
X_i = &\big\{\bar{r} \in [0, 1]^{<\omega} : \rng(\bar{r}) = \{c_{i,1}, \ldots, c_{i,m_i}\} 
\text{ and, for each $j = 1, \ldots, m_i$,} \\
&\text{there are between $(\alpha_{i,j} - \delta)|\bar{r}|$ and $(\alpha_{i,j} + \delta)|\bar{r}|$ coordinates in $\bar{r}$ } \\
&\text{which equal $c_{i,j}$} \big\}.
\end{align*}

\item For all $m_1, \ldots, m_k \in \mbbN^+$, $c_{i, j} \in [0, 1]$, $\alpha_{i, j} \in (0, 1]$, $i = 1, \ldots, k$
and $\varepsilon > 0$, there is $\delta > 0$ such that
if,  for $i = 1, \ldots, k$ and  $\bar{r}_i, \bar{\rho}_i \in [0, 1]^{<\omega}$, we have
\begin{enumerate}
\item $|\bar{\rho}_i| = |\bar{r}_i|$, 
\item $\mu_\infty^o(\bar{r}_i, \bar{\rho}_i) < \delta$, 
\item $\rng(\bar{r}_i) = \{c_{i,1}, \ldots, c_{i,m_i}\}$, and
\item for each $j = 1, \ldots, k_i$, there are between $(\alpha_{i,j} - \delta)|\bar{r}_i|$ and 
$(\alpha_{i,j} + \delta)|\bar{r}_i|$ coordinates in $\bar{r}_i$ which equal $c_{i, j}$,
\end{enumerate}
then $|F(\bar{r}_1, \ldots, \bar{r}_k) - F(\bar{\rho}_1, \ldots, \bar{\rho}_k)| < \varepsilon$.
\end{enumerate}
}\end{defin}

\noindent
We show that the two notions of admissibility 
(Definitions~\ref{alternative definition of admissible function}
and~\ref{definition of admissible function})
coincide:

\begin{prop}\label{equivalence of definitions of admissibility}
An aggregation function is admissible sensu novo if and only if it is admissible.
\end{prop}
 
\noindent
{\bf Proof.}
In order to make the notation less cluttered we only prove the proposition for ``unary''
aggregation functions $F : [0, 1]^{<\omega} \to [0, 1]$.
The generalization to $F : \big( [0, 1]^{<\omega}\big)^k \to [0, 1]$, for any $m > 1$, is essentially the same but we need to
refer to a sequence of $k$-tuples of sequences.

Let $F : [0, 1]^{<\omega} \to [0, 1]$ and suppose that $F$ is admissible.
We show that $F$ is admissible sensu novo.
We start with verifying Condition (1) of admissibility sensu novo.
Let $c_1, \ldots, c_k \in [0, 1]$ and $\alpha_1, \ldots, \alpha_k \in (0, 1]$.
Suppose that  $\delta > 0$ is small enough that for all $j$, $(\alpha_{j} - \delta) \in (0, 1)$ and
if $\alpha_j < 1$ then $(\alpha_{j} + \delta) \in (0,1)$. 
Let 
\begin{align*}
X = &\big\{\bar{r} \in [0, 1]^{<\omega} : \rng(\bar{r}) = \{c_{1}, \ldots, c_{k}\} 
\text{ and, for each $j = 1, \ldots, k$,} \\
&\text{there are between $(\alpha_{j} - \delta)|\bar{r}|$ and $(\alpha_{j} + \delta)|\bar{r}|$ coordinates in $\bar{r}$ } \\
&\text{which equal $c_{j}$} \big\}.
\end{align*}
Assume, towards a contradiction, that $F$ is not asymptotically uniformly continuous on $X$ with respect to $\mu_1^u$.
Then there is $\varepsilon > 0$ such that for all $\delta > 0$ and all $N \in \mbbN$ there are 
$\bar{r}_{\delta, N}$ and $\bar{\rho}_{\delta, N}$ in $X$ with 
$\mu_1^u ( \bar{r}_{\delta, N}, \bar{\rho}_{\delta, N}) < \delta$,
$|\bar{r}_{\delta, N}|, | \bar{\rho}_{\delta, N}| > N$ and 
$|F(\bar{r}_{\delta,N}) - F(\bar{\rho}_{\delta,N})| > \varepsilon$. 
We find convergence testing sequences as follows: 
Let $\bar{r}'_n = \bar{r}_{\delta, N}$ and $\bar{\rho}'_n = \bar{\rho}_{\delta, N}$,  
with $\delta = \frac{1}{n}$ and $N$ larger than the length of any  $\bar{r}_m$ and any $\bar{\rho}_m$ for $m < n$. 
Then $|\bar{r}'_n| < |\bar{r}'_{n+1}|$, $|\bar{\rho}'_n| < |\bar{\rho}'_{n+1}|$ 
and $|F(\bar{r}'_n) - F(\bar{\rho}'_n)| > \varepsilon$ for all $n$. 
Since $[0,1]^{k}$ is compact  the sequence of $k$-tuples of proportions 
\[
\bigg( \frac{| \{ i \leq |\bar{r}'_n| : r'_{n, i} = c_1 \} | }{|\bar{r}'_n|}, \ldots,
\frac{| \{ i \leq |\bar{r}'_n| : r'_{n, i} = c_k \} | }{|\bar{r}'_n|} \bigg)
\] 
where $r'_{n, i}$ is the $i$th coordinate of $\bar{r}'_n$,
has a convergent subsequence and the limit $(\alpha'_1, \ldots, \alpha'_k)$ lies in $(0,1)^k$ by the initial condition on $\delta$.
Without loss of generality, assume that this convergent subsequence is in fact the entire sequence. 
We claim that  $\bar{r}'_n$ and  $\bar{\rho}'_n$ are convergence testing sequences with parameters 
$c_1, \ldots, c_k$ and $\alpha'_1, \ldots, \alpha'_k$. 
Condition~(1) in the definition of convergence testing sequence
(Definition~\ref{definition of convergence testing})
is clear since  $\bar{r}'_n$ and  $\bar{\rho}'_n$ are strictly increasing in length as $n$ increases. 
Condition~(2) of the same definition is guaranteed for $\bar{r}'_n$ by the convergence of 
$\frac{\left| \{ i \leq |\bar{r}'_n| : \ r'_{n,i} = c_j \} \right| }{|\bar{r}'_n|}$  to $\alpha'_j$ (for each $j = 1, \ldots, k$).  
Since $\mu_1^u(\bar{r}'_n,\bar{\rho}'_n)$ converges to $0$, condition~(2) also holds for $\bar{\rho}'_n$.
Therefore $\bar{r}'_n$ and $\bar{\rho}'_n$ are convergence testing with the same parameters and thus by admissibility, 
$\underset{n \rightarrow \infty}{\lim}  |F(\bar{r}'_n) - F(\bar{\rho}'_n)| = 0$, {\em in contradiction to }
$|F(\bar{r}'_n) - F(\bar{\rho}'_n)| > \varepsilon$   for every $n$.

Now we show Condition~(2) of admissibility sensu novo. 
Let $\varepsilon > 0$. Fix $N \in \mbbN$. For $\delta < \frac{1}{n}$ any $\bar{r} \in [0, 1]^N$ is completely determined by 
Conditions~(2)(c)--(d) of admissibility sensu novo.    
Since, being admissible, $F$ is continuous at that 
$\bar{r} \in  [0, 1]^N$
there is a $\delta > 0$ such that for all $\bar{\rho} \in [0, 1]^N$,  whenever Conditions~2(b)--(d) of
admissibility sensu novo are satisfied,
$|F(\bar{r}) - F(\bar{\rho})| < \varepsilon$. 
Choose $\delta_N$ as the supremum of those $\delta$.
We need to show that there is a uniform lower bound $\delta_0 >0$ of all $\delta_N$ as $N$ ranges over $\mbbN$. 
Assume not. 
Then we can find a sequence of $N_n$ such that for all 
$n \in \mbbN$,  $N_{n+1} > N_n$ and $\delta_{N_n} < \frac{1}{n}$,
and we can find
$\bar{r}_n$ and $\bar{\rho}_n$ of length $N_n$ 
such that  $\bar{\rho}_n$  satisfies Conditions 2(a)--(d) of admissibility sensu novo
for $\delta = \frac{2}{n}$, but $|F(\bar{r}_n) - F(\bar{\rho}_n)| \geq \varepsilon$. 
We claim that both sequences are convergence testing with parameters    $c_j$ and $\alpha_j$. 
Indeed, the range of each $\bar{r}_n$ is  $\{c_1, \ldots, c_k\}$, and the proportion of entries equalling $c_j$  approaches $\alpha_j$,
as $n \to \infty$,
by Condition~2(d).
By Condition~2(b), for any open interval $I_j$ around $c_j$, there is an $N \in \mbbN$ such that for every $n > N$
an entry of $\bar{r}_n$ lies in $I_j$ if and only if the corresponding entry of $\rho_n$ lies in $I_j$. 
Therefore, the sequence of $\bar{\rho}_n$ is also convergence testing with parameters   $c_j$ and $\alpha_j$. 
Thus, as $F$ is admissible, $\underset{n \rightarrow \infty}{\lim}  |F(\bar{r}_n) - F(\bar{\rho}_n)| = 0$, 
{\em in contradiction to } $|F(\bar{r}_n) - F(\bar{\rho}_n)| \geq \varepsilon$   for every $n$.

Now we will proceed to show that if $F$ is admissible sensu novo then it is also admissible. 
So suppose that $F : [0, 1]^{<\omega} \to [0, 1]$ is admissible sensu novo.
We begin by verifying Condition~(1) of admissibility (Definition~\ref{alternative definition of admissible function}).
We need to show that $F$ is continuous on $[0, 1]^N$.
So let $\bar{r} \in [0, 1]^N$ and $\varepsilon > 0$.
Let $\rng(\bar{r}) = \{c_1, \ldots, c_k\}$ and let $\alpha_j$ be the proportion of entries in $\bar{r}$ that equal $c_j$.
Then by Condition~(2) of admissibility sensu novo
there is a $\delta > 0$ such that for all $\bar{\rho} \in [0, 1]^N$ with $\mu^o_\infty (\bar{r}, \bar{\rho}) < \delta$,
$|F(\bar{r}) - F(\bar{\rho})| < \varepsilon$. 
By the equivalence of norms in $\mbbR^N$, this suffices.

Now we show Condition~(2) of admissibility. 
Let $\bar{r}_n$ and $\bar{\rho}_n$, $n \in \mbbN$, be convergence testing with parameters 
$c_1, \ldots, c_k \in [0, 1]$ and $\alpha_1, \ldots, \alpha_k \in (0, 1]$. 
We show that 
$\underset{n \rightarrow \infty}{\lim}  |F(\bar{r}_n) - F(\bar{\rho}_n)| = 0$.
Let $I_1, \ldots I_k$ be disjoint open intervals such that $c_j \in I_j$.
As $\bar{r}_n$ is convergence testing there is $N$ such that if $n \geq N$, then 
$\rng(\bar{r}_n), \rng(\bar{\rho}_n) \subseteq \bigcup_{j=1}^k I_j$, and
\begin{equation}\label{properties of the two convergence testing sequences}
\underset{n\to\infty}{\lim} \frac{|\{i \leq |\bar{r}_n| : r_{n, i} \in I_j\}|}{|\bar{r}_n|} \ = \
\underset{n\to\infty}{\lim} \frac{|\{i \leq |\bar{\rho}_n| : \rho_{n, i} \in I_j\}|}{|\bar{\rho}_n|} \  = \ \alpha_j.
\end{equation}
Since we are only considering the limit, we can assume without loss of generality that $N =1$.  
Consider the sequences  $\bar{r}'_n$ and $\bar{\rho}'_n$ obtained by setting $r'_{n, l} =  c_j$ 
if  $r_{n, l} \in I_j$ (recall that different $I_j$ are disjoint),
and likewise $\rho'_{n, l} =  c_j$ if $\rho_{n, l} \in I_j$. 
By Condition~(1) of admissibility sensu novo, 
for every  $\varepsilon > 0$ there are $\delta > 0$ and $n_0$ depending only on $\varepsilon$ such that 
if $n > n_0$ and $\mu^u_1(\bar{r}'_n, \bar{\rho}'_n) < \delta$, then $|F(\bar{r}'_n) - F(\bar{\rho}'_n)| < \varepsilon$.
This together with~(\ref{properties of the two convergence testing sequences}) implies that
\[
\lim_{n\to\infty} |F(\bar{r}'_n) - F(\bar{\rho}'_n)| = 0.
\]
It now suffices to show that 
\[
\underset{n \rightarrow \infty}{\lim}  |F(\bar{r}_n) - F(\bar{r}'_n)| \ = \ 
\underset{n \rightarrow \infty}{\lim}  |F(\bar{\rho}_n) - F(\bar{\rho}'_n)| = 0.
\]
We claim that this is a consequence of Condition~(2) of admissibility sensu novo.  
We only show that the first limit equals 0, since the second limit is treated in the same way.
So let $\varepsilon > 0$ and choose an appropriate $\delta > 0$.
We need to show that, for all sufficiently large $n$, clauses (a)--(d) of Condition~(2) hold, with $\bar{r}'_n$ in the role of
$\bar{r}_i$ in condition~(2) of Definition~\ref{definition of admissible function} and $\bar{r}_n$ in the role of $\bar{\rho}_i$
in the same definition.
Clause~(a) is obvious.
Clause~(b) is true for sufficiently large $n$ since $\bar{r}_n$ is convergence testing, 
because we can just choose $I_j$ with diameter less than $\delta$.
Clause~(c) applied to $\bar{r}'_n$ is again clear by the definition of $\bar{r}'_n$.
Clause~(d) applied to $\bar{r}'_n$ is clear for sufficiently large $n$
because of~(\ref{properties of the two convergence testing sequences}).
This concludes the proof.
\hfill $\square$

\subsection{Asymptotic elimination of admissible aggregation functions}\label{Asymptotic elimination of admissible aggregation functions}

Throughout this section we assume that $\sigma$ is a finite and relational signature
and we let $\mbW_n$ be the set of all $\sigma$-structures with domain $[n]$.
Let $\mbbG$ be a lifted Bayesian network over $\sigma$ such that every aggregation formula
of $\mbbG$ is noncritical with respect to $\mbbG$.
Also let $\mbbP = (\mbbP_n : n \in \mbbN^+)$ be the sequence of probability distributions which is
induced by $\mbbG$.
Since the sequence of probability distributions $\mbbP$ is fixed throughout the section we will simply say that two
formulas are asymptotically equivalent when we mean that they are asymptotically equivalent with respect to
$\mbbP$.

When denoting an atomic $\sigma$-type by $p(\bar{x}, \bar{y})$, or a formula by $\varphi(\bar{x}, \bar{y})$, we assume that
$\rng(\bar{x}) \cap \rng(\bar{y}) = \es$.
For a first-order formula $\varphi(\bar{x}) \in PLA(\sigma)$, $n \in \mbbN^+$ and $\bar{a} \in [n]^{|\bar{x}|}$, we use the
notation
\[
\mbbP_n(\varphi(\bar{a})) = \mbbP_n\big(\{\mcA \in \mbW_n : \mcA \models \varphi(\bar{a})\}\big).
\]

In this section we prove that if $\varphi(\bar{x}) \in PLA(\sigma)$ and
all aggregation functions in $\varphi$ 
are admissible then there is a basic probability formula 
$\psi(\bar{x}) \in PLA(\sigma)$ such that 
$\varphi$ and $\psi$ are asymptotically equivalent; 
this is concluded by 
Corollary~\ref{elimination of aggregation functions if saturation conditions hold}
below and proves
Theorem~\ref{elimination of aggregation functions for an LBN}.

The proof of this result proceeds by induction on the complexity of $PLA(\sigma)$-formulas and we now outline the proof.
The main inductive step is to show that if $\varphi(\bar{x})$ denotes the formula
$F(\psi(\bar{x}, \bar{y}) : \bar{y} : p^=(\bar{x}, \bar{y}))$ where $F : [0,1]^{<\omega} \to [0, 1]^{<\omega}$ is 
an admissible aggregation function and 
$\psi$ is asymptotically equivalent to a basic probability formula, then $\varphi(\bar{x})$ is
asymptoticaly equivalent to a basic probability formula.
(The case for $F$ of higher arity than 1 is analogous.)
In fact, to begin with we will assume that $\psi(\bar{x}, \bar{y})$ is a basic probability formula 
and we will see that we can assume that it has the form
$\bigwedge_{i, j} (p_{i, j}(\bar{x}, \bar{y}) \to c_{i, j})$
 where each $p_{i, j}$ is (the conjunction of) a complete atomic $\sigma$-type,
$p_{i, j}\uhrc \bar{x} = p_{i, k} \uhrc \bar{x}$ for all $i, j$ and $k$, and
each $p_{i, j}$ implies $p^=$.

The crucial step of the proof is to analyse, 
for each $i$, each structure $\mcA \in \mbW_n$, and $\bar{a} \in [n]^{|\bar{x}|}$ such that $\bar{a}$
realizes the restriction of $p_{i, j}$ to $\bar{x}$, the sequence
\[
\bar{r} = \big(\mcA\big(\bigwedge_{i, j} ( p_{i, j}(\bar{a}, \bar{b}) \to c_i)\big) : \bar{b} \in [n]^{|\bar{y}|} \text{ and 
$p^=(\bar{a}, \bar{b})$ holds} \big).
\]
From Proposition~\ref{convergence and saturation if associated formulas are bpf}
below it follows that, for every $i$, with high probability, as $n \to \infty$, 
the proportion of $\bar{b} \in [n]^{|\bar{y}|}$, among those satisfying $p^=(\bar{a}, \bar{b})$, 
such that $\mcA \models p_{i, j}(\bar{a}, \bar{b})$ is close to some $\alpha_{i, j}$ which depends only on $p_{i, j}$ and $\mbbG$.
Therefore the proportion of $c_{i, j}$ in $\bar{r}$ is, with high probability, close to the sum of all
$\alpha_{i', j'}$ such that $c_{i', j'} = c_{i, j}$.
As $F$ is admissible, hence admissible sensu novo, it follows 
from condition~(1) in the definition of admissibility sensu novo
that $F(\bar{r})$ is, with high probability, close to a number $d_i$ which depends only on $F$,
$p_{i, j}$ (as $j$ ranges over its possible values) and $\mbbG$. 
Consequently, $\varphi(\bar{x})$ is asymptotically equivalent to a formula of the form 
$\bigwedge_i (q_i(\bar{x}) \to d_i)$ where $q_i(\bar{x}) = p_{i, j} \uhrc \bar{x}$.
This step is completed by Corollary~\ref{crucial step in eliminating F}.
Then we use this result and condition~(2) in the definition of admissibility sensu novo to show 
(in Proposition~\ref{elimination of one aggregation function})
that if $\psi(\bar{x}, \bar{y})$ is asymptotically equivalent to a 
basic probability formula (but is not necessarily itself a basic probability formula), then $\varphi(\bar{x})$
is asymptotically equivalent to a basic probability formula.

We first define the $\bar{y}$-dimension of an atomic $\sigma$ type $p(\bar{x}, \bar{y})$ which,  informally speaking,
is the number of degrees of freedom for the variables $\bar{y}$ once the variables $\bar{x}$ have been instantiated by
parameters from a structure.

\begin{defin}\label{definition of y-dimension}{\rm
Let $p(\bar{x}, \bar{y})$ be an atomic $\sigma$-type. The {\em $\bar{y}$-dimension of $p(\bar{x}, \bar{y})$},
denoted $\dim_{\bar{y}}(p)$, is the maximal $d \in \mbbN$ such that there are a $\sigma$-structure $\mcA$,
$\bar{a} \in A^{|\bar{x}|}$ and $\bar{b} \in A^{|\bar{y}|}$ such that $\mcA \models p(\bar{a}, \bar{b})$
and $|\rng(\bar{b}) \setminus \rng(\bar{a})| \geq d$.
}\end{defin}

\noindent
Let $p(\bar{x}, \bar{y})$ be an atomic $\sigma$-type and $d$ its $\bar{y}$-dimension.
We will, for large $n$, $\mcA \in \mbW_n$ and $\bar{a} \in [n]^{|\bar{x}|}$ that realizes $p \uhrc \bar{x}$
be interested in the proportion
\[
\frac{\big|\big\{\bar{b} \in [n]^{|\bar{y}|} : \mcA \models p(\bar{a}, \bar{b}) \big\}\big|}{n^d}.
\]
With the terminology of the next definition, the subsequent proposition tells that with high probability the above
proportion is close to a number which depends only on $p$, $p \uhrc \bar{x}$ and $\mbbG$. 
The same proposition also tells that for quantifier free formulas $\varphi(\bar{x})$, the probability that a tuple 
of parameters satisfies it converges, as $n \to \infty$, to a number that depends only on $\varphi$ and $\mbbG$.

\begin{defin}\label{definition of saturation}{\bf (Saturation and unsaturation)} {\rm
Let $\bar{x}$ and $\bar{y}$ be sequences of different variables such that 
$\rng(\bar{x}) \cap \rng(\bar{y}) = \es$ and let
$p(\bar{x}, \bar{y})$ and $q(\bar{x})$ be atomic $\sigma$-types such that $q \subseteq p$.
Let also $0 \leq \alpha \leq 1$ and $d = \dim_{\bar{y}}(p)$.
\begin{itemize}
\item[(a)] A finite $\sigma$-structure $\mcA$
is called {\em $(p, q, \alpha)$-saturated} if, whenever $\bar{a} \in A^{|\bar{x}|}$ and $\mcA \models q(\bar{a})$, then
$\big| \{\bar{b} \in A^{|\bar{y}|} : \mcA \models p(\bar{a}, \bar{b})\} \big| \geq \alpha |A|^d$.

\item[(b)] A finite $\sigma$-structure $\mcA$
is called {\em $(p, q, \alpha)$-unsaturated} if, whenever $\bar{a} \in A^{|\bar{x}|}$ and $\mcA \models q(\bar{a})$, then
$\big| \{\bar{b} \in A^{|\bar{y}|} : \mcA \models p(\bar{a}, \bar{b})\} \big| \leq \alpha |A|^d$.
\end{itemize}
}\end{defin}

\noindent
Note that if $p(\bar{x}, \bar{y})$ and $q(\bar{x})$ are as in the above definition
and the $\sigma$-structure $\mcA$ is $(p, q, 0)$-unsaturated, 
then $p(\bar{x}, \bar{y})$ is not realized in $\mcA$.
From \cite{Kop20} we can extract the following (with explanations that follow):

\begin{prop}\label{convergence and saturation if associated formulas are bpf}
(i) For every quantifier-free first-order formula $\varphi(\bar{x})$ over $\sigma$, 
$m \in \mbbN^+$ and every $\bar{a} \in [m]^{|\bar{x}|}$, $\lim_{n\to\infty} \mbbP_n(\varphi(\bar{a}))$ exists and
depends only on $\varphi(\bar{x})$ and $\mbbG$.
Moreover, the rate of convergence does not depend on $\bar{a}$, but only on $\varphi(\bar{x})$ and $\mbbG$.\\
(ii) Suppose that $p^=(\bar{x}, \bar{y})$ is a complete atomic $\es$-type such that $\dim_{\bar{y}}(p) > 0$, suppose that
$p(\bar{x}, \bar{y})$ is a
complete atomic $\sigma$-type such that $p^= \subseteq p$, and let $q(\bar{x}) = p \uhrc \bar{x}$.
Suppose that $\beta = \lim_{n\to\infty} \mbbP_n(p(\bar{a}, \bar{b}))$  
and $\gamma = \lim_{n\to\infty} \mbbP_n(q(\bar{a})) > 0$ where 
$\bar{a} \in [m]^{|\bar{x}|}$
and $\bar{b} \in [m]^{|\bar{y}|}$ are any tuples such that $p^=(\bar{a}, \bar{b})$ holds.
If $\alpha = \beta/\gamma$ then, for every $\varepsilon > 0$, 
\begin{align*}
&\lim_{n\to\infty} \mbbP_n\big( \{ \mcA \in \mbW_n : \text{ $\mcA$ is $(p, q, \alpha / (1 + \varepsilon))$-saturated and
$(p, q, \alpha (1 + \varepsilon))$-unsaturated} \} \big) \\ 
&= \ 1.
\end{align*}
(iii) The numbers $\beta$ and $\gamma$ from part~(ii) are products of numbers of the form
$\mu(R \ | \ \chi_{R,i})$ or $(1 - \mu(R \ | \ \chi_{R,i}))$ associated to $\mbbG$ as in part~(c) of
Definition~\ref{definition of BN}.
\end{prop}

\noindent
{\bf Proof.}
Part~(i) is a direct consequence of Theorem~3.15 in \cite{Kop20}, but we point out that there is an unfortunate typo in
the cited theorem where 
`$|\mbbP_n(\varphi(\bar{a})) - d| \leq 1 - e^{-cn}$' should read
`$|\mbbP_n(\varphi(\bar{a})) - d| \leq e^{-cn}$'.
Part~(ii) follows from Lemma~4.13 and Proposition~4.41 in \cite{Kop20} and induction on the maximal path rank
(\cite[Definition ~2.4]{Kop20}) of the
underlying DAG of the lifted Bayesian network $\mbbG$.
Part~(iii) follows from Lemma~4.12, Definition~4.18 and Corollary~4.19 in \cite{Kop20} and induction on the 
maximal path rank of the underlying DAG of $\mbbG'$.
\hfill $\square$
\\

\noindent
The next lemma states the expected fact that if $p_i(\bar{x}, \bar{y})$, $i = 1, \ldots, t$, is an enumeration without repetition of all 
complete atomic $\sigma$-types that extend a given complete atomic $\sigma$-type $q(\bar{x})$, 
then the sum, as $i = 1, \ldots, t$, of the numbers to which the probability of $p_i(\bar{x}, \bar{y})$ converges, 
conditioned on $q(\bar{x})$ being true, is 1.

\begin{lem}\label{the sum of the alphas is one}
Let $q(\bar{x})$ be a complete atomic $\sigma$-type, let
$p^=(\bar{x}, \bar{y})$ be a complete atomic $\es$-type which is consistent with $q(\bar{x})$,
 and let $p_1(\bar{x}, \bar{y}), \ldots, p_t(\bar{x}, \bar{y})$ enumerate,
without repetition, all complete atomic $\sigma$-types in the variables $\bar{x}\bar{y}$ which extend $q(\bar{x})$ and
$p^=(\bar{x}, \bar{y})$. 
Moreover, suppose that $\dim_{\bar{y}}(p^=) > 0$.
For $i = 1, \ldots, t$, let $\alpha_i = \beta_i/\gamma$ where 
$\beta_i = \lim_{n\to\infty} \mbbP_n(p_i(\bar{a}, \bar{b}))$, $\gamma = \lim_{n\to\infty} \mbbP_n(q(\bar{a}))$,
and $\bar{a} \in [m]^{|\bar{x}|}$ and $\bar{b} \in [m]^{|\bar{y}|}$ are chosen so that $p^=(\bar{a}, \bar{b})$ holds
and we assume that $\gamma > 0$.
Then $\alpha_1 + \ldots + \alpha_t = 1$.
\end{lem}

\noindent
{\bf Proof.}
Let $q(\bar{x})$, $p^=(\bar{x}, \bar{y})$, 
$p_1(\bar{x}, \bar{y}), \ldots, p_t(\bar{x}, \bar{y})$ and $\alpha_1, \ldots, \alpha_t$ be as assumed in the lemma.
Then, for all $n \geq m$ we have
\begin{align*}
1 &= \mbbP_n\bigg(\Big\{ \mcA \in \mbW_n : \mcA \models \bigvee_{i=1}^t p_i(\bar{a}, \bar{b}) \Big\} \ \Big| \ 
\Big\{ \mcA \in \mbW_n : \mcA \models q(\bar{a}) \text{ and } \mcA \models p^=(\bar{a}, \bar{b}) \Big\}\bigg) \\
&= \sum_{i=1}^t
\mbbP_n\bigg(\Big\{ \mcA \in \mbW_n : \mcA \models p_i(\bar{a}, \bar{b}) \Big\} \ \Big| \ 
\Big\{ \mcA \in \mbW_n : \mcA \models q(\bar{a}) \text{ and } \mcA \models p^=(\bar{a}, \bar{b}) \Big\}\bigg)  \\
&=  \sum_{i=1}^t \frac{\mbbP_n\big( \mcA \models p_i(\bar{a}, \bar{b}) \big)}{\mbbP_n\big( \mcA \models q(\bar{a}) \big)}
\end{align*}
and the last expression converges
to $\alpha_1 + \ldots + \alpha_t$ as $n \to \infty$. Therefore we must have $\alpha_1 + \ldots + \alpha_t = 1$.
\hfill $\square$
\\

\noindent
It will be convenient to argue in a context where, for some arbitrary $\delta > 0$ and $m \in \mbbN^+$,
we assume that if $p(\bar{x}, \bar{y})$ is a complete $\sigma$-type, $|\bar{x}| + |\bar{y}| \leq m$ and
$q(\bar{x}) = p \uhrc \bar{x}$, then all structures that we consider are 
$(p, q, \alpha/(1 + \delta))$-saturated and $(p, q, \alpha (1 + \delta))$-unsaturated
for some $\alpha$ that depends only on $p, q$ and $\mbbG$.
This is justified by the next definition and subsequent lemma.

\begin{defin}{\rm
For all $m, n \in \mbbN^+$ and $\delta > 0$, let $\mbY^{m, \delta}_n$ denote the set of all $\mcA \in \mbW_n$
such that for every complete atomic $\sigma$-type $p(\bar{x}, \bar{y})$ such that $|\bar{x}| + |\bar{y}| \leq m$,
if $q(\bar{x}) = p \uhrc \bar{x}$, $\dim_{\bar{y}}(p) > 0$ and $\alpha$ is the number given by 
Proposition~\ref{convergence and saturation if associated formulas are bpf}~(ii)
then $\mcA$ is $(p, q, \alpha/(1 + \delta))$-saturated and $(p, q, \alpha (1 + \delta))$-unsaturated.
}\end{defin}

\begin{lem}\label{Ydelta is almost sure}
For all $m \in \mbbN^+$ and $\delta > 0$, $\lim_{n\to\infty} \mbbP_n(\mbY^{m, \delta}_n) = 1$.
\end{lem}

\noindent
{\bf Proof.} 
Immediate from 
Proposition~\ref{convergence and saturation if associated formulas are bpf}~(ii),
since there are only finitely many complete atomic $\sigma$-types with at most $m$ variables.
\hfill $\square$

\begin{rem}\label{remark on higher arities}{\bf (Eliminating aggregation functions of higher arities)} {\rm
Lemmas~\ref{case of inconsistency with p-equality} --~\ref{crucial step in convergence of probability},
Corollary~\ref{crucial step in eliminating F}
and Proposition~\ref{elimination of one aggregation function}
below are stated and proved only for admissible aggregations functions $F : [0, 1]^{<\omega} \to [0, 1]$ but the results hold
also for admissible aggregation functions $F: \big([0, 1]^{<\omega}\big)^k \to [0, 1]$ where $k > 1$
and formulas $F(\psi_1(\bar{x}, \bar{y}), \ldots, \psi_k(\bar{x}, \bar{y}) : \bar{y} : p^=(\bar{x}, \bar{y}))$
where $\psi_1, \ldots, \psi_k$ are basic probability formulas. 
The proofs in the general case work out in essentially the same way but the notation becomes messier,
for example since the assumptions and notation introduced 
in Assumption~\ref{assumptions in results for eliminating an aggregation function}
for $\psi(\bar{x}, \bar{y})$ need to be considered for all $\psi_i(\bar{x}, \bar{y})$.
}\end{rem}

\noindent
We begin with a lemma which takes care of an odd case, which however is syntactically possible.

\begin{lem}\label{case of inconsistency with p-equality}
Let $p^=(\bar{x}, \bar{y})$ be a complete atomic $\es$-type and let, for $i = 1, \dots, k$, 
$c_i \in [0, 1]$ and let $p_i(\bar{x}, \bar{y})$ be an atomic $\sigma$-type which is {\em in}consistent with
$p^=(\bar{x}, \bar{y})$.
If $F : [0, 1]^{<\omega} \to [0, 1]$ is an admissible aggregation function then 
\[
F\big(\bigwedge_{i=1}^k (p_i(\bar{x}, \bar{y}) \to c_i) : \bar{y} : p^=(\bar{x}, \bar{y}) \big)
\]
is equivalent to a basic probability formula.
\end{lem}

\noindent
{\bf Proof.}
 Let $p'(\bar{x}) = p^= \uhrc \bar{x}$ and let 
$q_j(\bar{x})$, $j = 1, \ldots, l$, enumerate all complete atomic $\es$-types in the variables $\bar{x}$ 
which are different from $p'(\bar{x})$.

For all $m \in \mbbN^+$, let $\bar{r}_m$ be the constant 
sequence of length $m$ containing the number 1 in every entry.
As $F$ is admissible it is, by Proposition~\ref{equivalence of definitions of admissibility}, 
admissible sensu novo. For all $m$ and $m'$ we have $\mu_1^u(\bar{r}_m, \bar{r}_{m'}) = 0$ and
hence (by condition~(1) in Definition~\ref{definition of admissible function})
$F(\bar{r}_m) = F(\bar{r}_{m'})$. 
Let $c = F(\bar{r}_1)$.

Now it is straightforward to check, using the semantics of $PLA(\sigma)$
(Definition~\ref{semantics of PLA}), that
$F\big(\bigwedge_{i=1}^k (p_i(\bar{x}, \bar{y}) \to c_i) : \bar{y} : p^=(\bar{x}, \bar{y}) \big)$ 
is equivalent to 
$(p'(\bar{x}) \to c) \wedge \bigwedge_{j=1}^l(q_j(\bar{x}) \to 0)$.
\hfill $\square$
\\

\noindent
The next lemma justifies the making of some simplifying assumptions in the arguments that follow later
(see Assumption~\ref{assumptions in results for eliminating an aggregation function}).

\begin{lem}\label{removing redundant types}
Let $p^=(\bar{x}, \bar{y})$ be a complete atomic $\es$-type.
Let $p_i(\bar{x}, \bar{y})$, $i = 1, \ldots, k$, be atomic $\sigma$-types which are consistent with $p^=$
and let $p'_j(\bar{x}, \bar{y})$, $j = 1, \ldots, l$ be atomic $\sigma$-types which are not consistent with $p^=$.
Also let $c_1, \ldots, c_k, d_1, \ldots, d_l \in [0, 1]$.
For every aggregation function $F : [0, 1]^{<\omega} \to [0, 1]$,
the formula
\[
F\Big(\bigwedge_{i=1}^k (p_i(\bar{x}, \bar{y}) \to c_i) \wedge 
\bigwedge_{j=1}^l (p'_j(\bar{x}, \bar{y}) \to d_j) : \bar{y} : p^=(\bar{x}, \bar{y})\Big)
\]
is equivalent to
$F\big(\bigwedge_{i=1}^k (p_i(\bar{x}, \bar{y}) \to c_i)  : \bar{y} : p^=(\bar{x}, \bar{y})\big)$.
\end{lem}

\noindent
{\bf Proof.}
Let $\mcA$ be a finite $\sigma$-structure and let $\bar{a} \in A^{|\bar{x}|}$.
If $\bar{a}$ does not satisfy $p^= \uhrc \bar{x}$, then both formulas, with $\bar{x}$ interpreted as $\bar{a}$
have the value 0. 
If $\bar{a}$ satisfies $p^= \uhrc \bar{x}$, then 
\[
\mcA\Big(\bigwedge_{i=1}^k (p_i(\bar{a}, \bar{b}) \to c_i) \wedge 
\bigwedge_{j=1}^l (p'_j(\bar{a}, \bar{b}) \to d_j) \Big) \ = \
\mcA\Big(\bigwedge_{i=1}^k (p_i(\bar{a}, \bar{b}) \to c_i) \Big)
\]
for every $\bar{b} \in A^{|\bar{y}|}$ such that $p^=(\bar{a}, \bar{b})$ holds.
Hence $F$ is applied to the same sequence in both cases and therefore both formulas in the statement of
the lemma get the same value.
\hfill $\square$
\\

\noindent
The previous two lemmas justify the addition of the following assumptions in the main part of the proof
of the asymptotic elimination of aggregation functions.

\begin{assump}\label{assumptions in results for eliminating an aggregation function} {\rm
In Lemma~\ref{crucial step in convergence of probability}
and Corollary~\ref{crucial step in eliminating F}
we make the following assumptions:
Let $\kappa \in \mbbN^+$ and let $\bar{x}$ and $\bar{y}$ be sequences of distinct variables such that 
$\rng(\bar{x}) \cap \rng(\bar{y}) = \es$ and $|\bar{x}| + |\bar{y}| \leq \kappa \in \mbbN^+$.
Let $p^=(\bar{x}, \bar{y})$ be a complete atomic $\es$-type, let $l = \dim_{\bar{y}}(p^=)$ and let
$\psi(\bar{x}, \bar{y})$ denote the basic probability formula
\[
\bigwedge_{i=1}^s\bigwedge_{j=1}^{t_i} \big(p_{i,j}(\bar{x}, \bar{y}) \rightarrow c_{i,j}\big),
\]
where we may, without loss of generality, assume
that each $p_{i,j}(\bar{x}, \bar{y})$ is a complete atomic $\sigma$-type and $p^= \subseteq p_{i, j}$.
Furthermore,
we assume (by reordering if necessary) that for all $i = 1, \ldots, s$ and all $1 \leq j, j' \leq t_i$, $p_{i, j} \uhrc \bar{x} = p_{i, j'} \uhrc \bar{x}$.
Let $q_i(\bar{x}) = p_{i,1} \uhrc \bar{x}$ for each $i$.
Without loss of generality we may also assume that the $p_{i,j}(\bar{x}, \bar{y})$, 
$i = 1, \ldots, s$, $j = 1, \ldots, t_i$, enumerate all complete atomic 
$\sigma$-types with free variables $\bar{x}, \bar{y}$ which extend $p^=(\bar{x}, \bar{y})$
(because $c_{i, j}$ is allowed to be zero).
Note that $l = \dim_{\bar{y}}(p_{i, j})$ for all $i$ and $j$.
}\end{assump}

\noindent
We are now ready for the main technical lemma, the proof of which uses 
condition~(1) of Definition~\ref{definition of admissible function} of admissibility sensu novo.

\begin{lem}\label{crucial step in convergence of probability}
Suppose that $F : [0, 1]^{<\omega} \to [0, 1]$ is an admissible aggregation function.
Fix an index $1 \leq i \leq s$.
Then there is $d_i \in [0, 1]$, depending only on $q_i$ and $F$, such that
for every $\varepsilon > 0$ there is $\delta > 0$ such that for all sufficiently large $n$, all $\mcA \in \mbY_n^{\kappa, \delta}$, and all $\bar{a} \in [n]^{|\bar{x}|}$,
if $\mcA \models q_i(\bar{a})$, then
\[
\big| \mcA\big(F\big( \psi(\bar{a}, \bar{y}) : \bar{y} : p^=(\bar{a}, \bar{y})\big)\big) - d_i \big| < \varepsilon.
\]
\end{lem}

\noindent
{\bf Proof.}
Let $\varepsilon > 0$.
The conclusion of the lemma will follow if we can show that there is $\delta > 0$ such that for all sufficiently large $n_1$ and $n_2$,
all $\mcA_1 \in \mbY_{n_1}^{\kappa, \delta}$, all $\mcA_2 \in \mbY_{n_2}^{\kappa, \delta}$, all $\bar{a}_1 \in[n_1]^{|\bar{x}|}$ and all
$\bar{a}_2 \in [n_2]^{|\bar{x}|}$, if $\mcA_1 \models q_i(\bar{a}_1)$ and $\mcA_2 \models q_i(\bar{a}_2)$, then
\begin{equation}\label{convergence in the most basic case}
\Big| \mcA_1\big(F\big(\psi(\bar{a}_1, \bar{y}) : \bar{y} : p^=(\bar{a}_1, \bar{y})\big)\big) - 
\mcA_2\big(F\big(\psi(\bar{a}_2, \bar{y}) : \bar{y} : p^=(\bar{a}_2, \bar{y})\big)\big) \Big| < \varepsilon.
\end{equation}
Towards the end of the argument we will see that the assumption that $F$ is admissible implies that such $\delta$ exists.

Let $\delta > 0$ and suppose that 
$\mcA_1 \in \mbY_{n_1}^{\kappa, \delta}$, $\mcA_2 \in \mbY_{n_2}^{\kappa, \delta}$, $\bar{a}_1 \in[n_1]^{|\bar{x}|}$, 
$\bar{a}_2 \in [n_2]^{|\bar{x}|}$, $\mcA_1 \models q_i(\bar{a}_1)$ and $\mcA_2 \models q_i(\bar{a}_2)$.
Recall that $l = \dim_{\bar{y}}(p^=) = \dim_{\bar{y}}(p_{i, j})$ for all $j$.

For $k = 1, 2$ let 
\[
\bar{r}_k = \big(\mcA_k\big(\psi(\bar{a}_k, \bar{b})\big) : 
\bar{b} \in [n_k]^{|\bar{y}|} \text{ and $p^=(\bar{a}, \bar{b})$ holds}\big)
\]
and observe that, for every $\bar{b} \in [n_k]^{|\bar{y}|}$,
\[
\mcA_k\big(\psi(\bar{a}_k, \bar{b})\big) = c_{i, j} \ \text{ if } \
\mcA_k \models p_{i,j}(\bar{a}_k, \bar{b}).
\]
It follows that for each $k = 1,2$ and every $j = 1, \ldots, t_i$, 
every $\bar{b} \in [n_k]^{|\bar{y}|}$ such that $\mcA_k \models p_{i,j}(\bar{a}_k, \bar{b})$ contributes to 
a coordinate $c_{i,j}$ in the sequence $\bar{r}_k$.

If $l = 0$ then  $p^=$ has a unique extension to a complete atomic $\sigma$-type with 
variables $\bar{x}, \bar{y}$ and which includes $q_i(\bar{x})$,
so $t_i = 1$ and, for $k = 1, 2$,  $p_{i, 1}(\bar{a}_k, \bar{y})$ is realized by the unique tuple which realizes 
$p^=(\bar{a}_k, \bar{y})$.
Hence $|\bar{r}_1| = | \bar{r}_2| = 1$ and for $k = 1, 2$ the unique entry of $\bar{r}_k$ is $c_{i, 1}$,
so $\bar{r}_1 = \bar{r}_2$ and therefore $\mu_1^u(\bar{r}_1, \bar{r}_2) = 0$.

Now suppose that $l > 0$, so 
Proposition~\ref{convergence and saturation if associated formulas are bpf}~(ii)
is applicable.
Let $\alpha_j$ be the number associated to $p_{i, j}$ by
Proposition~\ref{convergence and saturation if associated formulas are bpf}~(ii).
Since $\mcA_k \in \mbY^{\kappa, \delta}_{n_k}$ for $k = 1, 2$ it follows that
\begin{align}\label{saturation property expressed by upper and lower bounds}
\frac{\alpha_j}{(1 + \delta)}(n_k)^l \leq 
\big|\big\{ \bar{b} \in [n_k]^{|\bar{y}|} : \mcA_k \models p_{i,j}(\bar{a}_k, \bar{b}) \big\}\big| 
\leq \alpha_j(1 + \delta)(n_k)^l.
\end{align}

Suppose that $c \in [0, 1]$ and that there are exactly $m$ indices $j = j_1, \dots, j_m$ such that $c_{i, j} = c$.
It follows from~(\ref{saturation property expressed by upper and lower bounds}) that, for each $k = 1, 2$ and
sufficiently large $n_k$
the number $c$ will occur between 
\[
(\alpha_{j_1} + \ldots + \alpha_{j_m}) (n_k)^l /(1 + \delta) \ \  \text{ and } \ \
(\alpha_{j_1} + \ldots + \alpha_{j_m}) (n_k)^l(1 + \delta)
\]
times in $\bar{r}_k$.
In particular, if all $\alpha_{i_1}, \ldots, \alpha_{j_m}$ are 0, then $c_{i, j}$ does not occur in $\bar{r}_k$.
From Lemma~\ref{the sum of the alphas is one} we get $\alpha_1 + \ldots + \alpha_{t_i} = 1$.
It now follows from definitions~\ref{definition of associated function} and~\ref{definition of the metric} that
$\mu_1^u(\bar{r}_1, \bar{r}_2) \leq \delta g(t_i)$
where $g(t_i)$ depends only on $t_i$.

We assume that $F$ is  admissible and hence it is admissible sensu novo,
by Proposition~\ref{equivalence of definitions of admissibility}.
From Condition~(1) of the definition of admissibility sensu novo
(Definition~\ref{definition of admissible function})
it follows that if $\delta > 0$ is small enough 
and $n_1$ and $n_2$ large enough, then $|F(\bar{r}_1) - F(\bar{r}_2)| < \varepsilon$
and hence
\[
\Big| \mcA_1\big(F\big(\psi(\bar{a}_1, \bar{y}) : \bar{y} : p^=(\bar{a}_1, \bar{y})\big)\big) - 
\mcA_2\big(F\big(\psi(\bar{a}_2, \bar{y}) : \bar{y} : p^=(\bar{a}_2, \bar{y})\big)\big) \Big| < \varepsilon.
\]
\hfill $\square$

\begin{rem}\label{remark on necessity of p-equality}{\rm 
Suppose for a moment that we would allow formulas of the form 
$F(\psi(\bar{x}, \bar{y}) : \bar{y})$, where $F$ is an aggregation function, with the semantic interpretation
$\mcA\big(F(\psi(\bar{a}, \bar{y}) : \bar{y})\big) = 
F\big(\mcA\big(\psi(\bar{a}, \bar{b})\big) : \bar{b} \in A^{|\bar{y}|}\big)$.
Then condition~(1) of Definition~\ref{definition of admissible function}
of admissibility sensu novo which was used in the proof of
Lemma~\ref{crucial step in convergence of probability}
is no longer, in general, applicable in the same proof.

To exemplify this, suppose that $\sigma$ is the empty signature, that $\bar{x}$ is the empty sequence of variables
and that $\bar{y} = (y_1, y_2)$.
Consider the formula `$(y_1 = y_2) \to 1/2$' which we denote by $\psi(\bar{x}, \bar{y})$ to use the
same notation as in the proof of
Lemma~\ref{crucial step in convergence of probability}.
Let $\mcA_1$, $\mcA_2$, $\bar{a}_1$ and $\bar{a}_2$ be as in the proof of
Lemma~\ref{crucial step in convergence of probability} (so $\bar{a}_1$ and $\bar{a}_2$ are empty in this example)
and for $k = 1, 2$ let 
\[
\bar{r}_k = \big(\mcA_k\big(\psi(\bar{a}_k, \bar{b})\big) : 
\bar{b} \in [n_k]^{|\bar{y}|} \big).
\]
Then, for $k = 1, 2$, exactly $n_k$ entries of  $\bar{r}_k$ will be $1/2$ and exactly $n_k(n_k - 1)$ entries of $\bar{r}_k$
will be 1, so the proportion of `$1/2$' is $1/(n_k - 1)$ which is not zero but tends to zero as $n_k$ tends to infinity.
It follows that $\mu_1^u(\bar{r}_1, \bar{r}_2) \to 0$ as $n_1, n_2 \to \infty$.
Since the parameters denoted $\alpha_{i, j}$ in condition~(1) of the definition of admissibility sensu novo are required to be
nonzero we cannot use condition~(1) to conclude that $F(\bar{r}_1, \bar{r}_2)$ is as small as we like if 
$\mu_1^u(\bar{r}_1, \bar{r}_2)$ is sufficiently small.
}\end{rem}

\begin{cor}\label{crucial step in eliminating F}
Suppose that
$F : [0, 1]^{<\omega} \to [0, 1]$ is an admissible aggregation function.
Then there is a basic probability formula $\theta(\bar{x})$ such that
for every $\varepsilon > 0$ there is $\delta > 0$ such that for all sufficiently large $n$, 
all $\mcA \in \mbY_n^{\kappa, \delta}$, 
and all $\bar{a} \in [n]^{|\bar{x}|}$, we have
\[
\big| \mcA\big(F\big(\psi(\bar{a}, \bar{y}) : \bar{y} : p^=(\bar{a}, \bar{y})\big)\big) \ - \  
\mcA\big(\theta(\bar{a}) \big) \big| < \varepsilon.
\]
\end{cor}

\noindent
{\bf Proof.}
Recall that $q_i(\bar{x})$ are assumed to be as in 
Assumption~\ref{assumptions in results for eliminating an aggregation function}, 
so each $q_i(\bar{x})$ is consistent with the restriction of $p^=$ to the variables $\bar{x}$.
For every $i = 1, \ldots, s$, let $d_i \in [0, 1]$ be as in 
Lemma~\ref{crucial step in convergence of probability}.
Let $q'_1(\bar{x}), \ldots, q'_m(\bar{x})$ enumerate all complete atomic $\es$-types in 
the variables $\bar{x}$ which are different from $p^= \uhrc \bar{x}$.
We show that we can let $\theta(\bar{x})$ be the formula
$\bigwedge_{i=1}^s (q_i(\bar{x}) \to d_i) \ \wedge \ 
\bigwedge_{j=1}^m (q'_j(\bar{x}) \to 0)$.
Let $\varepsilon > 0$.
Let $\mcA \in  \mbY_n^{\kappa, \delta}$ and $\bar{a} \in [n]^{|\bar{x}|}$.

If $\mcA \models q'_j(\bar{a})$ for some $j$, then (no matter what $\delta$ is)
\[
\mcA\big(F\big(\psi(\bar{a}, \bar{y}) : \bar{y} : p^=(\bar{a}, \bar{y})\big)\big) \ = \ 
0 \ = \ \mcA\Big(\bigwedge_{i=1}^s (q_i(\bar{a}) \to d_i) \ \wedge \ 
\bigwedge_{j=1}^m (q'_j(\bar{a}) \to 0) \Big).
\]

Now suppose that $\bar{a}$ satisfies $p^=(\bar{x})$ and hence it satisfies $q_i(\bar{x})$ for some $i$.
Then 
\[
\mcA\Big( \bigwedge_{i=1}^s (q_i(\bar{x}) \to d_i) \ \wedge \ 
\bigwedge_{j=1}^m (q'_j(\bar{x}) \to 0) \Big) \ = \  d_i.
\]
It follows from Lemma~\ref{crucial step in convergence of probability}
that if $\delta > 0$ is small enough, then for every $i = 1, \ldots, s$, all sufficiently large $n$, all $\mcA \in \mbY_n^{\kappa, \delta}$, 
and all $\bar{a} \in [n]^{|\bar{x}|}$,
if $\mcA \models q_i(\bar{a})$, then
\[
\big| \mcA\big(F\big( \psi(\bar{a}, \bar{y}) : \bar{y} : p^=(\bar{a}, \bar{y})\big)\big) - d_i \big| < \varepsilon.
\]
Consequently
\[
\Big| \mcA\big(F\big(\psi(\bar{a}, \bar{y}) : \bar{y} : p^=(\bar{a}, \bar{y})\big)\big) \ - \ 
\mcA\Big(\bigwedge_{i=1}^s (q_i(\bar{a}) \rightarrow d_i)
\ \wedge \ 
\bigwedge_{j=1}^m (q'_j(\bar{x}) \to 0) \Big) \Big| < \varepsilon.
\]
\hfill $\square$

\noindent
The next proposition states that one admissible aggregation function can be asymptotically eliminated
and this is the main step in the inductive proof of 
Corollary~\ref{elimination of aggregation functions if saturation conditions hold}
The proof of the proposition uses condition~(2) of 
Definition~\ref{definition of admissible function} 
of admissibility sensu novo.

\begin{prop}\label{elimination of one aggregation function}
Suppose that $\varphi(\bar{x}, \bar{y}), \psi(\bar{x}, \bar{y}) \in PLA(\sigma)$ are asymptotically equivalent formulas and that
$\psi(\bar{x}, \bar{y})$ is a basic probability formula.
Also suppose that $p^=(\bar{x}, \bar{y})$ is a complete atomic $\es$-type which is consistent with each one of $\psi(\bar{x}, \bar{y})$
and $\varphi(\bar{x}, \bar{y})$.
If $F : [0, 1]^{<\omega} \to [0, 1]$ is an admissible aggregation function,
then $F\big(\varphi(\bar{x}, \bar{y}) : \bar{y} : p^=(\bar{x}, \bar{y})\big)$ is asymptotically equivalent to a basic probability formula.
\end{prop}

\noindent
{\bf Proof.}
Let $F : [0, 1]^{<\omega} \to [0, 1]$ be an admissible aggregation function.
Suppose that $\varphi(\bar{x}, \bar{y}), \psi(\bar{x}, \bar{y}) \in PLA(\sigma)$ are asymptotically equivalent and that
$\psi(\bar{x}, \bar{y})$ is a basic probability formula.
Let $\kappa = |\bar{x}| + |\bar{y}|$ and $\varepsilon > 0$.
By Lemmas~\ref{case of inconsistency with p-equality}
and~\ref{removing redundant types}
and Corollary~\ref{crucial step in eliminating F}
there is a basic probability formula $\theta(\bar{x})$ such that 
for all small enough $\delta > 0$ and large enough $n$, if $\mcA \in \mbY_n^{\kappa, \delta}$ 
and $\bar{a} \in [n]^{|\bar{x}|}$, then
\begin{equation}\label{the conclusion of a previous corollary}
\big| \mcA\big(F\big(\psi(\bar{a}, \bar{y}) : \bar{y} : p^=(\bar{a}, \bar{y})\big)\big) \ - \ 
\mcA\big(\theta(\bar{a})\big) \big| < \varepsilon/2.
\end{equation}

For any real $\delta > 0$ and any $n \in \mbbN^+$ let 
\begin{align*}
\mbX_n^\delta = \big\{ \mcA \in \mbW_n : &\text{ for all $\bar{a} \in [n]^{|\bar{x}|}$ and all $\bar{b} \in [n]^{|\bar{y}|}$
such that $p^=(\bar{a}, \bar{b})$ holds},\\
&\big|\mcA(\varphi(\bar{a}, \bar{b})) - \mcA(\psi(\bar{a}, \bar{b}))\big| < \delta \big\}.
\end{align*}
Since $\varphi(\bar{x}, \bar{y})$ and $\psi(\bar{x}, \bar{y})$ are asymptotically equivalent we have
$\lim_{n\to\infty}\mbbP_n(\mbX_n^\delta) =~1$.
By Lemma~\ref{Ydelta is almost sure}
we also have $\lim_{n\to\infty}\mbbP_n(\mbY_n^{\kappa, \delta}) = 1$, so it follows that
$\lim_{n\to\infty}\mbbP_n(\mbX_n^\delta \cap \mbY_n^{\kappa, \delta}) =~1$.

It now suffices to prove that if $\delta > 0$ is small enough, then for all sufficiently large $n$,
all $\mcA \in \mbX_n^\delta \cap \mbY_n^{\kappa, \delta}$ and all $\bar{a} \in [n]^{|\bar{x}|}$ we have
\begin{equation}\label{to be proved in eliminating F in the general case}
\big| \mcA\big(F\big(\varphi(\bar{a}, \bar{y}) : \bar{y} : p^=(\bar{a}, \bar{y})\big)\big) - 
\mcA\big(\theta(\bar{a})\big) \big| < \varepsilon.
\end{equation}
Statement~(\ref{to be proved in eliminating F in the general case}) 
follows from~(\ref{the conclusion of a previous corollary})
and the following (to be proved)
\begin{equation}\label{difference between F(varphi) and F(psi)}
\big| \mcA\big(F\big(\varphi(\bar{a}, \bar{y}) : \bar{y} : p^=(\bar{a}, \bar{y})\big)\big) - 
\mcA\big(F\big(\psi(\bar{a}, \bar{y}) : \bar{y} : p^=(\bar{a}, \bar{y})\big)\big) < \varepsilon/2.
\end{equation}
Hence it remains to prove that if $\delta > 0$ is small enough 
then~(\ref{difference between F(varphi) and F(psi)}) holds
for all sufficiently large $n$,
all $\mcA \in \mbX_n^\delta \cap \mbY_n^{\kappa, \delta}$ and all $\bar{a} \in [n]^{|\bar{x}|}$.

Let $\mcA \in \mbX_n^\delta \cap \mbY_n^{\kappa, \delta}$ and $\bar{a} \in [n]^{|\bar{x}|}$.
If $\bar{a}$ does not satisfy $p^= \uhrc \bar{x}$ then 
\[
\mcA\big(F\big(\varphi(\bar{a}, \bar{y}) : \bar{y} : p^=(\bar{a}, \bar{y})\big)\big) \ = \ 0 \ = \ 
\mcA\big(F\big(\psi(\bar{a}, \bar{y}) : \bar{y} : p^=(\bar{a}, \bar{y})\big)\big).
\]
Now suppose that $\bar{a}$ satisfies $p^= \uhrc \bar{x}$. Then the following two sequences
are nonempty:
\begin{align*}
&\bar{r} = \big(\mcA\big(\varphi(\bar{a}, \bar{b})\big) : 
\bar{b} \in [n]^{|\bar{y}|} \text{ and $p^=(\bar{a}, \bar{b})$ holds} \big), \\
&\bar{\rho} = \big(\mcA\big(\psi(\bar{a}, \bar{b})\big) : 
\bar{b} \in [n]^{|\bar{y}|} \text{ and $p^=(\bar{a}, \bar{b})$ holds} \big).
\end{align*}
First suppose that $\psi(\bar{x}, \bar{y})$ has the form $\bigwedge_{i=1}^k (p_i(\bar{x}, \bar{y}) \to c_i)$
where each $p_i$ is an atomic $\sigma$-type which is {\em in}consistent with $p^=$. 
Then all entries in $\bar{\rho}$ are equal to 1.
Since $\mcA \in \mbX_n^\delta$ we get $\mu_\infty^o(\bar{r}, \bar{\rho}) < \delta$.
Since $F$ is admissible, hence admissible sensu novo, if follows 
from Condition~(2) of the definition of admissibility sensu novo,
that if $\delta$ is small enough, then $|F(\bar{r}) - F(\bar{\rho})| < \varepsilon/2$
and~(\ref{difference between F(varphi) and F(psi)}) follows immediately from this.

If $\psi$ does not have the form just considered, then, using
Lemma~\ref{removing redundant types},
we may assume that $\psi$ satisfies 
Assumption~\ref{assumptions in results for eliminating an aggregation function}.
Then we can argue in the same way as we argued in the proof of 
Lemma~\ref{crucial step in convergence of probability} and conclude 
that there are numbers $t \in \mbbN^+$, $c_1, \ldots, c_t \in [0, 1]$ and $\alpha_1, \ldots, \alpha_t \in (0, 1]$,
depending only on $\psi$ and $\mbbG$,
such that $\rng(\bar{\rho}) = \{c_1, \ldots, c_t\}$ and, for $c \in [0, 1]$, if 
$c_{i_1}, \ldots, c_{i_m}$ enumerates all $c_i$ such that $c_i = c$, then
$c$ appears  between 
\[
(\alpha_{i_1} + \ldots + \alpha_{i_m}) n^l / (1 + \delta) \ \ \text{ and } \ \
(\alpha_{i_1} + \ldots + \alpha_{i_m}) n^l(1 + \delta)
\]
times in $\bar{\rho}$.
Since $F$ is admissible, hence admissible sensu novo, if follows 
from Condition~(2) of the definition of admissibility sensu novo,
that if $\delta$ is small enough, then $|F(\bar{r}) - F(\bar{\rho})| < \varepsilon/2$
which implies that~(\ref{difference between F(varphi) and F(psi)}) holds.
\hfill $\square$

\begin{cor}\label{elimination of aggregation functions if saturation conditions hold}
Let $\varphi(\bar{x}) \in PLA(\sigma)$ and suppose that
all aggregation functions in $\varphi$ are admissible.
Then $\varphi(\bar{x})$ is asymptotically equivalent to a basic probability formula.
\end{cor}

\noindent
{\bf Proof.}
We use induction on the complexity of formulas.
If the aggregation rank is 0, that is, if the formula is aggregation-free, then the conclusion follows from
Lemma~\ref{quantifier-free formulas are equivalent to bpf}, since equivalence implies asymptotic equivalence.

Suppose that the aggregation rank of $\varphi(\bar{x})$ is larger than 0.
We have one case for each way in which $\varphi$ can be constructed from simpler formulas,
as in parts~(4) and~(5) of Definition~\ref{syntax of PLA}.
We start with part~(4), the ``propositional constructions'', and consider only one of the subcases, since the other are treated in the same way.
Suppose that $\varphi(\bar{x})$ is the formula $\psi(\bar{x}) \wedge \chi(\bar{x})$.
By the induction hypothesis, there are basic probability formulas $\psi'(\bar{x})$ and $\chi'(\bar{x})$ such that $\psi(\bar{x})$ and
$\psi'(\bar{x})$ are asymptotically equivalent and $\chi(\bar{x})$ and $\chi'(\bar{x})$ are asymptotically equivalent.
By Lemma~\ref{preservation of asymptotic equivalence under connectives},
$\psi(\bar{x}) \wedge \chi(\bar{x})$ is asymptotically equivalent to $\psi'(\bar{x}) \wedge \chi'(\bar{x})$.
The formula $\psi'(\bar{x}) \wedge \chi'(\bar{x})$ is aggregation-free, hence 
(by Lemma~\ref{quantifier-free formulas are equivalent to bpf})
it is equivalent to a basic probability formula $\varphi'(\bar{x})$.
Then $\varphi(\bar{x})$ and $\varphi'(\bar{x})$ are asymptotically equivalent.

Now we turn to part~(5) of Definition~\ref{syntax of PLA} and suppose that $\varphi(\bar{x})$ has the form 
$F(\psi_1(\bar{x}, \bar{y}), \ldots, \psi_k(\bar{x}, \bar{y}) : \bar{y} : p^=(\bar{x}, \bar{y}))$ where $F$ denotes
an admissible aggregation function.
Then each $\psi_i$ is simpler than $\varphi$ so each $\psi_i(\bar{x}, \bar{y})$ is, by the induction hypothesis,
asymptotically equivalent to a basic probability formula.
Then Proposition~\ref{elimination of one aggregation function}
combined with 
Remark~\ref{remark on higher arities}
implies that $\varphi(\bar{x})$ is asymptotically equivalent to a basic probability formula.
\hfill $\square$

\begin{rem}\label{Computing an asymptotically equivalent formula without aggregation functions}
{\bf (Computing an asymptotically equivalent formula without aggregation functions)}
{\rm 
Corollary~\ref{elimination of aggregation functions if saturation conditions hold}
guarantees that for every $\varphi(\bar{x}) \in PLA(\sigma)$ with only admissible aggregation functions there is a 
basic probability formula $\psi(\bar{x})$ which is asymptotically equivalent to $\varphi(\bar{x})$.
If $\varphi$ is aggregation-free then 
Lemma~\ref{quantifier-free formulas are equivalent to bpf}
guarantees the existence of such $\psi$
and in practice such $\psi(\bar{x})$ can be 
constructed by, for every complete atomic $\sigma$-type $p(\bar{x})$, computing the value $c_p$ that $\varphi(\bar{x})$ takes
if $p(\bar{x})$ is satisfied. Then $\psi(\bar{x})$ will be (up to equivalence) the conjunction of 
formulas of the form $p(\bar{x}) \to c_p$.
(If $\varphi$ is aggregation-free and without free variables, then $\varphi$ has the same value in all structures and
we compute this value, call it $c_p$, and then $\varphi$ is equivalent to the basic probability formula $\top \to c_p$.)

If $\varphi$ is not aggregation-free we first reduce the problem to finding, for each subformula
of $\varphi(\bar{x})$, say $\varphi'(\bar{x})$,
a basic probability formula which is asymptotically equivalent to $\varphi'(\bar{x})$.
Assuming this has been done and (which is the nontrivial case) that
$\varphi(\bar{x})$ has the form $F(\psi_1(\bar{x}, \bar{y}), \ldots, \psi_k(\bar{x}, \bar{y}) : \bar{y} : p^=(\bar{x}, \bar{y}))$,
where $F$ is admissible,
we proceed like this, where to simplify notation we assume that $k = 1$. 
Thus let $\varphi(\bar{x})$ be $F(\psi(\bar{x}, \bar{y}) : \bar{y} : p^=(\bar{x}, \bar{y}))$.
So by assumption we have computed a basic probability formula $\psi'(\bar{x}, \bar{y})$
which is asymptotically equivalent to $\psi(\bar{x}, \bar{y})$.
By modifying $\psi'$ if necessary we can assume that it has the form
\[
\bigwedge_{i=1}^s\bigwedge_{j=1}^{t_i} \big(p_{i,j}(\bar{x}, \bar{y}) \rightarrow c_{i,j}\big)
\]
where reach $p_{i, j}$ is a complete atomic $\sigma$-type and for all $i, j, k$, $p_{i, j} \uhrc \bar{x} = p_{i, k} \uhrc \bar{x}$.

If every $p_{i, j}$ is inconsistent with $p^=$ then the proof of
Lemma~\ref{case of inconsistency with p-equality}
shows how to form a basic probability formula which is equivalent to $\varphi(\bar{x})$.
Otherwise, we may (justified by Lemma~\ref{removing redundant types})
remove all $p_{i, j}$ which are inconsistent with $p^=$ and
assume that that all conditions in 
Assumption~\ref{assumptions in results for eliminating an aggregation function} hold.

According to Proposition~\ref{convergence and saturation if associated formulas are bpf},
for each $p_{i, j}$, all $\bar{a} \in [m]^{|\bar{x}|}$ and $\bar{b} \in [m]^{|\bar{y}|}$ such that $p^=(\bar{a}, \bar{b})$ holds,
the limits $\lim_{n\to\infty} \mbbP_n(p_{i, j}(\bar{a}, \bar{b}))$ and $\lim_{n\to\infty} q_i(\bar{b})$ exist,
where $q_i = p_{i, j} \uhrc \bar{x}$, and are products of numbers associated to $\mbbG$
(that is, numbers denoted $\mu(R \ | \ \chi_{R, i})$ in 
Definition~\ref{definition of BN}). 
If these limits are denoted $\beta_{i, j}$ and $\gamma_i$, respectively, then let $\alpha_{i, j} = \beta_{i, j}/\gamma_i$.

The next task is, for each $i$, to find the limit of $F(\bar{r})$, as the length of $\bar{r}$ tends to infinity and
$\bar{r}$ has the properties of $\bar{r}_1$ (or $\bar{r}_2$) in the proof of 
Lemma~\ref{crucial step in convergence of probability}, with $\alpha_{i, j}$ abbreviated as $\alpha_j$.
More precisely, given some small $\delta > 0$, large $n$ and assuming that $l = \dim_{\bar{y}}(p^=) > 0$, we construct $\bar{r}$
of length $n$ as follows: if $c \in [0, 1]$ and
there there are exactly $m$ indices $j = j_1, \dots, j_m$ such that $c_{i, j} = c$, 
then we let $\bar{r}$ have between 
$(\alpha_{j_1} + \ldots + \alpha_{j_m}) n^l /(1 + \delta)$ and 
$(\alpha_{j_1} + \ldots + \alpha_{j_m}) n^l(1 + \delta)$ occurrences of $c$
(and if $c \neq c_{i, j}$ for all $j$, then $\bar{r}$ has no occurrence of $c$).
Since $F$ is assumed to be admissible, hence admissible sensu novo,
the limit of $F(\bar{r})$ for such $\bar{r}$ as its length tends to infinity exists
and let us suppose that the limit is $d_i$ (for each index $i$).
Then $F(\psi'(\bar{x}, \bar{y}) : \bar{y} : p^=(\bar{x}, \bar{y}))$ is asymptotically equivalent
to $\bigwedge_{i=1}^s (q_i(\bar{x}) \to d_i)$, where $q_i = p_{i, j}\uhrc \bar{x}$, as implied by
Corollary~\ref{crucial step in eliminating F} (with $\psi'$ in place of $\psi$).
Proposition~\ref{elimination of one aggregation function}
implies that $F(\psi(\bar{x}, \bar{y}) : \bar{y} : p^=(\bar{x}, \bar{y}))$ is asymptotically equivalent to 
$\bigwedge_{i=1}^s (q_i(\bar{x}) \to d_i)$.

Although we know that  the limit of $F(\bar{r})$, for $\bar{r}$ as described above, exists as the length of $\bar{r}$ tends to infinity,
it may not be clear how to compute it. 
In this case we can still estimate the limit (assuming that $F(\bar{r})$ can be estimated
with arbitrarily high precision for every relevant $\bar{r}$), 
by choosing large $n$, constructing $\bar{r}$ as above and computing
(or estimating) $F(\bar{r})$.
Since $F$ is admissible we know that for any $\varepsilon > 0$, if $n$ is large enough and $\delta$ small enough,
then $F(\bar{r})$ is within distance $\varepsilon$ of the limit, by
Condition~(1) of Definition~\ref{definition of admissible function}.
}\end{rem}

\section{Conclusion}\label{Conclusion}

\noindent
We have considered what we call {\em probability logic with aggregation functions (PLA)} for expressing queries. 
$PLA$ uses aggregation functions instead of quantifiers, but can express all queries that are expressible in first-order logic
by using the aggregation functions max and min.
The motivation comes from data mining, machine learning and statistical relational artificial intelligence where 
aggregation over a domain is often done with aggregation functions, for example the arithmetic mean of a sequence of reals.
Since the mean of a sequence need not be 0 or 1, even if all entries in the sequence are 0 or 1, $PLA$ is a many valued
logic with values in the unit interval $[0, 1]$.
A typical query in this context is to ask ``Is the value of (a sentence) $\varphi$ in the interval $I$?''.

Then our aim was to study the asymptotic behaviour, as the domain size tends to infinity, of the probability of a query expressible with $PLA$
with respect to certain probability distributions of relevance withing statistical relational AI.
As there are so many different kinds of aggregation functions, we do not expect to find a single result
that covers the asymptotic behaviour of $PLA$-formulas with arbitrary aggregation functions.
Hence we identified what we call {\em admissible} 
(or intuitively ``partially uniformly continuous'') aggregation functions for which we could prove asymptotic results.
The arithmetic and geometric means and max and min are admissible, but we also gave examples of several other admissible
aggregation functions.
We demonstrated the expressive power of $PLA$ restricted to admissible aggregation functions by, for example, showing that
every stage in the approximation of the SimRank can be expressed by a $PLA$-formula with only admissible aggregation functions
(and by similar but simpler arguments one can show that every approximation stage of the Page rank 
\cite{BP} can be expressed by a $PLA$-formula
with only admissible aggregation functions).

We have used the formalism {\em lifted Bayesian network} for inducing, for any finite relational signature $\sigma$,
a probability distribution on the set of $\sigma$-structures with a given finite domain.
Roughly speaking, a lifted Bayesian network for $\sigma$ is a directed acyclic graph with vertex set $\sigma$ which
specifies (conditional) probabilities to each $R \in \sigma$ by case distinctions expressed by 
formulas of {\em conditional probability logic (CPL)}, that use only the parents of $R$ in the directed acyclic graph.
$CPL$ is a 2-valued logic that extends first-order logic and with which one can express that a relative frequency (of events expressed by
$CPL$-formulas) belongs to a given interval, or that the difference between two relative frequencies belongs to a certain interval.
This type of construction in $CPL$ can be iterated as many times as one likes, just as quantifiers can be nested in first-order logic.

With this set up our main result was
that every $PLA$-formula $\varphi(\bar{x})$ with only admissible aggregation functions is
asymptotically equivalent to a $PLA$-formula $\psi(\bar{x})$ {\em without} aggregation functions, which in rough terms means that
the values of the two formulas will with high probability be almost the same (and $\psi(\bar{x})$
can only take finitely many different values).
From the proof one can extract a procedure for finding such $\psi$ and the procedure needs only $\varphi$ and the 
lifted Bayesian network as input.

From the main result we derive a convergence law for $PLA$-formulas with only admissible aggregation functions.
It states that for any such formula $\varphi(\bar{x})$ 
there are $\alpha_1, \ldots, \alpha_k, c_1, \ldots, c_k \in [0, 1]$ (for some $k$) such that
the sum of the $\alpha_i$ is 1 and,
for any sequence of parameters $\bar{a}$ from the domain, every $\varepsilon > 0$ and $i$,
with probability tending to $\alpha_i$ the value of  $\varphi(\bar{a})$ will belong to $[c_i - \varepsilon, c_i + \varepsilon]$.

The studies begun here have continued in \cite{KW2} where we, among other things, prove similar results
in a context allowing more  probability distributions,
including such where, with high probability, some or all relations are ``sparse'',
but at the cost of only allowing what we call {\em strongly admissible} aggregation functions in $PLA$-formulas.
The arithmetic and geometric means are strongly admissible but max and min are not.
Due to results about random graphs \cite{SS} it is impossile, in general, to asymptotically eliminate max and min
from $PLA$-formulas in the context of sparse graphs.

\end{document}